\documentclass[10pt,aps,prd,nofootinbib,superscriptaddress,twocolumn,preprintnumbers,balancelastpage]{revtex4-2}

\usepackage[colorlinks=true
,urlcolor=blue
,anchorcolor=blue
,citecolor=blue
,filecolor=blue
,linkcolor=blue
,menucolor=blue
,linktocpage=true
,pdfproducer=medialab
,pdfa=true
]{hyperref}
\usepackage{amssymb,amsmath,multirow,slashed,cleveref,comment} 

\usepackage{graphicx}
\usepackage[dvipsnames]{xcolor}
\usepackage{booktabs}
\usepackage{lipsum}
\usepackage{bm}
\usepackage{setspace}
\usepackage[T1]{fontenc}
\usepackage{lettrine}
\usepackage{Zallman}

\makeatletter
\renewcommand{\@seccntformat}[1]{%
  \ifcsname prefix@#1\endcsname
    \csname prefix@#1\endcsname
  \else
    \csname the#1\endcsname\quad
  \fi}

\newcommand{\KCL}{\affiliation{Department of Physics, King’s College London, Strand, London, WC2R 2LS, UK}}

\begin{document}

\preprint{ \includegraphics[width=0.4\textwidth]{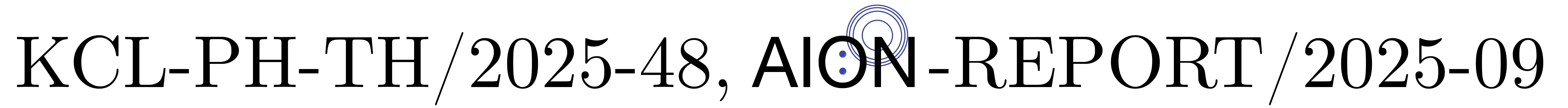}  }

\title{Searching for screened scalar forces with long-baseline atom interferometers}

\author{Hannah Banks}
\email{hannah.banks@nyu.edu}
\affiliation{DAMTP, University of Cambridge, Wilberforce Road, Cambridge, CB3 0WA, UK}
\affiliation{Center for Cosmology and Particle Physics, Department of Physics, New York University, New York, NY 10003, USA}
\author{John Carlton}
\email{john.carlton@kcl.ac.uk}
\KCL
\affiliation{Department of Physics and Astronomy, University of Kentucky, Lexington, KY, 40506-0055, USA}

\author{Benjamin Elder}
\email{b.elder@imperial.ac.uk}
\affiliation{Abdus Salam Centre for Theoretical Physics, Imperial College London, Prince Consort Road, London SW7 2AZ, UK}

\author{Thomas Hird}
\email{t.m.hird@bham.ac.uk}
\affiliation{Clarendon Laboratory, Department of Physics, University of Oxford, Parks Road, OX1 3PU, UK}
\affiliation{School of Physics and Astronomy, University of Birmingham, Edgbaston, Birmingham B15 2TT, UK}

\author{Christopher McCabe}
\email{christopher.mccabe@kcl.ac.uk}
\KCL

\begin{abstract}
\begin{center}
{\bf Abstract}
\vspace{-7pt}
\end{center}

Screened scalars are ubiquitous in many dark-sector models. 
They give rise to non-trivial fifth forces whilst evading experimental constraints through density-dependent screening mechanisms. 
We propose equipping a 10\,m-scale long-baseline atom interferometer with an annular planar source mass inside the vacuum chamber to search for such screened fifth forces. 
Two key challenges arise: distinguishing the static fifth force from backgrounds, and isolating it from the plate's Newtonian gravity. We introduce the `$Q$-flip protocol', which alternates between interferometry sequences to induce controllable time-dependence, aiding signal extraction and de-trending of transient noise.
We further develop an \emph{in situ} calibration procedure to characterise the plate's Newtonian gravity and reach shot-noise-limited sensitivity.
We show that our proposal could test theoretically motivated parameter space, advancing existing bounds in chameleon and symmetron screened scalar models by $1$ to $1.5$ orders of magnitude. 
Our proposal is directly applicable to forthcoming experiments, such as AION-10 or VLBAI, and is readily extensible to broader theoretical models and longer baselines.


\end{abstract}
\maketitle
\interfootnotelinepenalty=10000

\section{Introduction}
\lettrine{M}{ultiple} cosmological observations indicate that the vast majority of the energy density of the Universe is harboured by some as-yet unknown entities: dark matter and dark energy. Despite being the subject of a multi-faceted experimental campaign, which has been continually rejuvenated over several decades with the emergence of new technologies, we are yet to uncover any direct evidence for either substance. Determining the nature and properties of the Universe's dark sector remains one of the most pressing open questions in contemporary physics. 

Many of the theoretical constructions proposed to address these conundrums invoke new light scalar bosons with couplings to the Standard Model (SM). The exchange of these fields between SM particles generically induces new long-range monopole-monopole fifth forces in normal matter.  Such phenomena have been subject to extensive scrutiny over a range of scales by a variety of experiments
searching for modifications to the gravitational inverse square law \cite{Adelberger:2003zx, Burrage:2016bwy,Brax:2021wcv}.  If the couplings to the SM matter fields are non-universal, the induced force is composition-dependent and additionally subject to stringent constraints from experiments such as the MICROSCOPE satellite~\cite{Berge:2017ovy, MICROSCOPE:2022doy} and E{\"o}t-Wash torsion balance experiments~\cite{Kapner:2006si,Wagner:2012ui} which are sensitive to violations of the Weak Equivalence Principle.

The most minimal and most readily studied theoretical target in scalar fifth-force tests arises from the exchange of a single scalar boson, yielding the familiar Yukawa potential~\cite{Yukawa:1935xg}. More exotic force profiles can however be generated by more complex particle exchange (see~\cite{Banks:2020gpu} for a fully generic treatment). A further class of models that has achieved considerable attention are those in which the new light degree of freedom is subject to a  \textit{screening mechanism} where the effective coupling to matter depends on the ambient matter density. In regions of high density, such as the environments in which laboratory and astrophysical measurements are made, non-linear terms in the scalar equation of motion become large, resulting in an effective decoupling of the new field from the SM.  

In this way, these theories allow for scalar couplings to matter of strength comparable to, or stronger than gravity whilst remaining phenomenologically viable~\cite{Joyce:2014kja}.  Such theories, many of which have received attention as possible dark energy candidates, exhibit parametrically different behaviour on laboratory scales compared to Yukawa-like forces, demanding carefully tailored detection strategies.  Some of the most popular models belonging to this class include the chameleon~\cite{Khoury:2003aq,Khoury:2003rn} and the symmetron~\cite{Hinterbichler:2010es,Hinterbichler:2011ca}, which form the theoretical targets for this work.  We refer to theories with a screening mechanism as `screened scalars'.

Atom interferometers measure the phase difference between spatially delocalised quantum superpositions of clouds of atoms~\cite{Abend:2020djo}, and are highly sensitive to acceleration~\cite{McGuirk:2002zz}. This makes them well-placed to probe new forces that manifest as a modification to gravity. Indeed, table-top scale atom interferometry has already been successfully used to probe screened fifth forces~\cite{Burrage:2014oza, Hamilton:2015zga, Elder:2016yxm, Burrage:2016rkv, Sabulsky:2018jma}, with optical lattice interferometry~\cite{Panda:2023nir} currently providing the leading bounds on chameleon theories over large swathes of parameter space. 
There have since been several additional proposals to search for chameleons using atom interferometry, including an optical-lattice setup~\cite{PhysRevD.109.123515} and the DESIRE project in the Einstein Elevator~\cite{garcion2025darkenergysearchatom}.

A new generation of long-baseline atom interferometers is emerging following the development of novel techniques and protocols. These are configured as gradiometers, comprising two or more atom interferometers interrogated by common laser pulses~\cite{snadden1998measurement}. Several prototypes of such configurations  at the $\mathcal{O}(10)$\,m and $\mathcal{O}(100)$\,m scale are planned or under construction, including AION~\cite{Badurina:2019hst,Bongs:2025rqe}, AICE~\cite{Baynham:2025pzm}, ELGAR~\cite{Canuel:2020cxb}, MAGIS-100~\cite{MAGIS-100:2021etm}, MIGA~\cite{Canuel:2017rrp}, VLBAI~\cite{Hartwig:2015iza}, and ZAIGA~\cite{Zhan:2019quq}. Together, these efforts form the foundation of the global TVLBAI initiative~\cite{TVLBAI:2024,Abdalla:2024sst}, which aims to establish instruments with even longer (km-scale) baselines.

Despite numerous studies investigating the potential of long-baseline atom interferometers to probe feebly interacting phenomena such as gravitational waves (GW)~\cite{Dimopoulos:2007cj, Dimopoulos:2008sv, Yu:2010ss, Graham:2012sy, Chaibi:2016dze, Graham:2017pmn, Loriani:2018qej, Schubert:2019ycf, Baum:2023rwc, Badurina:2024rpp, Schach:2025jmp,Schaffrath:2025rsn, Sala:2025uqh} and ultra-light dark matter (ULDM)~\cite{Graham:2015ifn, Geraci:2016fva, Arvanitaki:2016fyj, Badurina:2021lwr, Badurina:2021rgt, DiPumpo:2022muv,Badurina:2023wpk, DiPumpo:2023hfk,Derr:2023bqk, Banks:2024sli, Blas:2025qrw,Badurina:2025xwl},
their application to screened scalar fifth forces has received less attention.\footnote{The possibility of probing \textit{unscreened} spin-dependent (monopole-dipole) forces with such instruments was explored in Ref.~\cite{Abe:2024idx}.} 
We address this gap by proposing an experimental protocol tailored to 10\,m-scale vertical baseline gradiometers which have been equipped with an internal annular source mass positioned at the top of the vacuum chamber.\footnote{While these instruments could, in principle, probe any scalar monopole-monopole fifth force, the protocol and set-up proposed here is specifically tailored for screened fifth forces. Yukawa-like forces would likely require different design choices.}

Two challenges arise with this configuration: the fifth force from the fixed plate is static, making it difficult to distinguish from other static backgrounds, and the plate's Newtonian gravity dominates over the scalar signal. We introduce complementary protocols to address each. First, the `$Q$-flip protocol' which alternates between interferometry sequences with different sensitivities to the source mass, creating a time-modulated signal from the static fifth force without any mechanical motion of the source mass. Second, we exploit the different distance dependences of the scalar force and Newtonian gravity: the scalar force drops off exponentially beyond a characteristic range whilst gravity follows an inverse square law. By measuring the gradiometer response at different atom-plate separations, we characterise the gravitational contribution and subtract it to isolate the signal from screened scalars.  We show that a 10\,m instrument could improve existing chameleon and symmetron bounds by more than an order-of-magnitude, decisively testing theoretically motivated parameter space.

The outline of this paper is as follows. We begin in~Sec.~\ref{sec:model} by outlining the chameleon and symmetron models. In Sec.~\ref{sec:setup} we detail the experimental configuration, measurement protocols, and numerical calculation of the scalar field profile in the vicinity of an annular source mass, and its contribution to the gradiometer phase, a calculation not previously performed. In Sec.~\ref{sec:projections} we provide sensitivity forecasts for the chameleon and symmetron models. Systematic effects are explored in Sec.~\ref{sec:systematics}, and we conclude in Sec.~\ref{sec:conclusions}. Appendix~\ref{app:boundary-conditions} contains details of the numerical approximations used to calculate the scalar field profile.

{\small \bf Conventions:} In this work we use the mostly-plus metric convention, typically work in natural units in which $c = \hbar = 1$, and define the reduced Planck mass as $M_\mathrm{Pl} = (8 \pi G_N)^{-1/2} \approx 2 \times 10^{18}~\mathrm{GeV}$, where $G_N$ is the Newtonian constant of gravitation.

\section{Screened Scalar Models}\label{sec:model}

The non-linearities in the scalar potential that result in screening can generically arise in one of two ways: either as a result of a self-interaction potential or due to the presence of derivative interactions. The former class of models, which includes both the chameleon and the symmetron, yields rich phenomenology on laboratory scales and will be the theoretical targets in this work.  It is likely that theories in the latter class, which includes galileons and disformally coupled scalars, could also be analogously probed with our experimental protocol.\footnote{See \cite{Brax:2011sv} for a discussion of galileon phenomenology in the context of laboratory experiments.} See~\cite{Joyce:2014kja,Burrage:2016bwy,CANTATA:2021asi,Brax:2021wcv} for a classification of all screening mechanisms, and various constraints on their associated models.

The models we consider in this paper involve a real scalar field $\varphi$ with a canonical kinetic term described by the action 
\begin{equation}
    S_\varphi = \int {\rm d}^4x\,\sqrt{-g}\left(-\frac{1}{2}(\partial\varphi)^2-V(\varphi)-A(\varphi)\rho_{\rm m}\right)~,
\end{equation}
where $V(\varphi)$ is the self-interaction potential of the scalar field, $\rho_\mathrm{m}$ is the local energy density of SM matter fields which are assumed to be non-relativistic, and $A(\varphi)$ gives the coupling of the scalar to the SM fields.  Through this coupling, $\rho_\mathrm{m}$ sources the scalar field.\footnote{We do not consider quantum mechanical corrections to the scalar couplings~\cite{Upadhye:2012vh}. For the symmetron mechanism, quantum
corrections can result in up to a $30\%$ weaker fifth force than the classical prediction alone~\cite{Udemba:2025csd} (depending on the precise scenario).  We follow standard convention in our forecasts and consider just classical predictions.}
In general the scalar field obeys a wave equation; however, provided the source term $\rho_\mathrm{m}$ is time-independent, as is the case in this work, the field can be assumed to be static. The equation of motion for the scalar field in a static environment is
\begin{equation}
    \vec\nabla^2\varphi = \frac{{\rm d}V}{{\rm d}\varphi}+\frac{{\rm d}A}{{\rm d}\varphi}\rho_{\rm m}~.
    \label{scalar-eom}
\end{equation}

An extended object in this scalar field experiences a classical force which contributes acceleration 
\begin{equation}
    \vec a_\varphi= - \lambda_a\frac{{\rm d}A}{{\rm d}\varphi}\vec\nabla\varphi(\vec x)~,
    \label{eq:acc}
\end{equation}
where $\lambda_a \in (0, 1]$ is a model-dependent `screening factor' that quantifies the strength of the interaction between the object and the scalar field.  The screening factor is derived by solving Eq.~\eqref{scalar-eom} in the vicinity of the extended object, and in general depends on the ambient field value (see e.g.~\cite{Hui:2009kc,Brax:2022olf} for a derivation).

In chameleon or symmetron theories, extended objects that are sufficiently small and/or underdense have $\lambda_a \approx 1$, and are said to be `unscreened'.  Conversely, objects that are very large and/or dense will have $\lambda_a < 1$ and are said to be `screened', thus experiencing a reduced acceleration. In general, this is a result of non-linear terms in the scalar equation of motion becoming significant in the vicinity of the source matter, due to the field perturbation generated by the extended object. 

We now introduce the two specific models that will form the theoretical targets of this study.  They are distinguished by their self-interaction potential and matter coupling, and exhibit two different classes of screening mechanism.

\subsection{Chameleon}

The chameleon is a scalar field that is epitomised by an environment-dependent mass that becomes large in dense environments~\cite{Khoury:2003aq,Khoury:2003rn}.  {Intuitively, deep inside a dense object, the chameleon mass becomes so large that the range of interaction is smaller than the size of the object.  As such, only matter in a `thin shell' near the surface of the object contributes to external interactions.}  In the typical realisation of this mechanism, the self-interaction potential and matter coupling are
\begin{equation}
    V_{\rm ch}(\varphi)=\frac{\Lambda^5}{\varphi}~,\quad A_{\rm ch}(\varphi)=\frac{\varphi}{M}~,
    \label{eq:ch_potentials}
\end{equation}
where $M, \Lambda$ are energy scales characterising the interaction strengths.
This model is a common target for laboratory searches for screened scalar forces and we thus adopt it as one of our two benchmark models.  

The region of parameter space of greatest theoretical interest in this model has $\Lambda$ around the dark energy scale (i.e.,~$\Lambda_\mathrm{DE} = 2.4~\mathrm{meV}$) and $M \sim M_\mathrm{Pl}$, which results in a force on unscreened objects that is comparable to ordinary gravity. While these parameters are particularly theoretically motivated, the broader parameter space remains of interest. Our sensitivity projections therefore span several orders of magnitude in both $\Lambda$ and $M$.

The chameleon screening factor for a spherical object of radius $R_\mathrm{obj}$ and density $\rho_\mathrm{obj}$ is~\cite{Khoury:2003aq,Khoury:2003rn,Sabulsky:2018jma}
\begin{equation}
    \lambda_{\rm a,ch}\approx {\rm min}\left(\frac{3 M \varphi_{\rm env}}{\rho_{\rm obj}R^2_{\rm obj}}, 1\right)~,
    \label{eq:chameleon-screening-factor}
\end{equation}
where $\varphi_\mathrm{env}$ is the ambient field value in the vicinity of the spherical object, i.e., the field that would be present in the absence of the object itself.

To give some intuition, let us examine Eq.~\eqref{eq:chameleon-screening-factor}  using a simplified analytical estimate for an atom inside a vacuum chamber (in practice, we solve the field equation numerically for the full experimental geometry). The atom's mass is dominated by its nucleus, with mass $m_\mathrm{obj} \approx A ~\mathrm{GeV}$ and radius $R_{\rm obj}\simeq 1.25 \times A^{1/3} ~{\rm fm}$ where $A$ is the atomic mass number, for which we take $A = 87$ for $^{87}$Sr atoms. 
The density can be computed from these quantities in the usual way, $\rho_{\rm obj} \simeq m_\mathrm{obj}/\left(\frac{4}{3}\pi R_{\rm obj}^3\right)$. Finally, we must determine the ambient field value inside the chamber. 
 For an infinitely long cylindrical vacuum chamber of inner radius $R_\mathrm{vac}$, the field at the centre is $\varphi_{\rm env} = \xi (2 \Lambda^5 R_\mathrm{vac}^2)^{1/3}$ where $\xi\approx0.68$ (see Appendix~\ref{app:boundary-conditions}).
Taking $R_\mathrm{vac} = 5~\mathrm{cm}$, we find
\begin{equation}
    \lambda_\mathrm{a,ch}^{\rm atom} = \min \left( 10^8 \left( \frac{\Lambda}{~\mathrm{meV}} \right)^{5/3} \left( \frac{M}{M_\mathrm{Pl}} \right), 1 \right)~.
    \label{chameleon-sf-atom}
\end{equation}
In contrast, for a macroscopic object, e.g., a marble with $R_\mathrm{obj} \approx \mathrm{cm}$ and $\rho \approx \mathrm{g/cm}^3$, the screening factor is $\sim10^{-2}$.
We thus see that whilst macroscopic objects are generally screened, atoms are not (over a wide range of parameter space). This is a key property that enables chameleon theories to be tested with atom interferometry.

\subsection{Symmetron}

The symmetron model~\cite{Hinterbichler:2010es} exhibits a second archetype of screening known as the symmetron mechanism. This is characterised by a potential and matter coupling that cause the scalar field to decouple from ordinary matter in dense environments.  Although the screening mechanism is distinct from that of the chameleon, both result in similar qualitative behaviour in that the fifth force is suppressed in dense environments.  

In a typical realisation of this mechanism the symmetron potential and matter coupling are
\begin{equation}
    V_{\rm sym}(\varphi) = -\frac{1}{2}\mu^2\varphi^2+\frac{\lambda}{4}\varphi^4~,\quad A_{\rm sym}(\varphi) = \frac{\varphi^2}{2M^2}~,
    \label{eq:sym_potentials}
\end{equation}
where $\mu$ is a mass scale, $\lambda$ is a dimensionless coupling, and $M$ controls the strength of the matter coupling.
This model results in a critical density $\rho_\mathrm{crit} \equiv \mu^2 M^2$.  In regions where the matter field density $\rho_\mathrm{m}$ is below $\rho_\mathrm{crit}$, the field tends towards a vacuum expectation value (VEV) $v \equiv \mu / \sqrt{\lambda}$.  If the ambient density is above $\rho_\mathrm{crit}$, the field is instead driven towards~$0$.  Whilst this specific realisation exhibits the symmetron mechanism by design, this mechanism can also emerge from more general theories via radiative corrections~\cite{Burrage:2016xzz}.

Symmetron screening may be intuitively understood as follows: as described above, the field transitions from its VEV in low-density regions to zero in high-density regions. This transition begins within a thin shell near the surface of dense objects.  Once the field reaches zero inside the object, it remains constant throughout the interior. As a result, the external field gradient is sourced only by matter in this thin shell near the surface, just as with the chameleon, resulting in a screened external force.   

The symmetron model has several free parameters. A common benchmark sets $\mu$ near the dark energy scale ($\mu \sim \Lambda_\mathrm{DE}$) with $M\sim\mathrm{TeV}$ and $\lambda\sim1$ such that $v / M^2 \sim M_\mathrm{Pl}^{-1}$, resulting in a fifth force on unscreened objects comparable to Newtonian gravity. However, as with the chameleon, the broader parameter space remains of interest. In this work, we demonstrate sensitivity across a wide range of parameters spanning many orders of magnitude.

The screening factor for an extended spherical object in this model is~\cite{Hinterbichler:2010es, Sabulsky:2018jma}
\begin{equation}
    \lambda_{\rm a,sym}\approx {\rm min}\left(\frac{M^2}{\rho_{\rm obj}R^2_{\rm obj}},1\right)~.
    \label{eq:sf_sym}
\end{equation}
Considering the idealised setup of $^{87}$Sr atoms in an empty chamber where the field reaches its VEV, the screening factor is
\begin{equation}
    \lambda_\mathrm{a, sym}^{\rm atom} = \min \left( 1.3 \left( \frac{M}{\mathrm{GeV}} \right)^2 , 1\right)~.
    \label{symm-sf-atom}
\end{equation}
By comparison, for a macroscopic marble ($R_\mathrm{obj} \approx \mathrm{cm},\, \rho \approx \mathrm{g/cm}^3$) the pre-factor is $\sim10^{-10}$.  As in the chameleon case, whilst macroscopic objects are generally screened, atoms are not over a wide range of parameter space.

In what follows, we use these theoretical details to compute the scalar fifth force on atoms in free-fall inside an atom interferometer vacuum chamber.  The local matter density from the source mass and the vacuum chamber walls determines the profile of the scalar field $\varphi$, according to Eq.~\eqref{scalar-eom}. The atoms are treated as extended spherical objects with screening factors given by Eqs.~\eqref{eq:chameleon-screening-factor} and~\eqref{eq:sf_sym} for the chameleon and symmetron, respectively.

\section{Experimental Set-up and Measurement Strategy}\label{sec:setup}

Screened scalar models can be probed through precision measurements of anomalous forces. 
As evident from Eq.~\eqref{eq:acc}, two key factors determine the force: the screening factor $\lambda_a$ and the field gradient $\vec \nabla \varphi$.
Atom interferometers are exceptionally sensitive accelerometers while also being  
well-suited to optimise both factors. 
Firstly, the vacuum chamber provides a low ambient matter density, allowing the scalar field to develop large values such that atoms remain unscreened $(\lambda_a \approx 1)$ over much of the parameter space of interest. 
Secondly, a source mass can be positioned inside the vacuum chamber, supplying a large field gradient whilst avoiding the strong screening from the chamber walls that would occur if the source was positioned externally.  The non-trivial environmental dependence of screened scalars gives rise to  specific design considerations and novel measurement protocols, which we now describe.

\subsection{Experimental configuration}

\begin{figure}[!t]
    \centering
    \includegraphics[width=\linewidth]{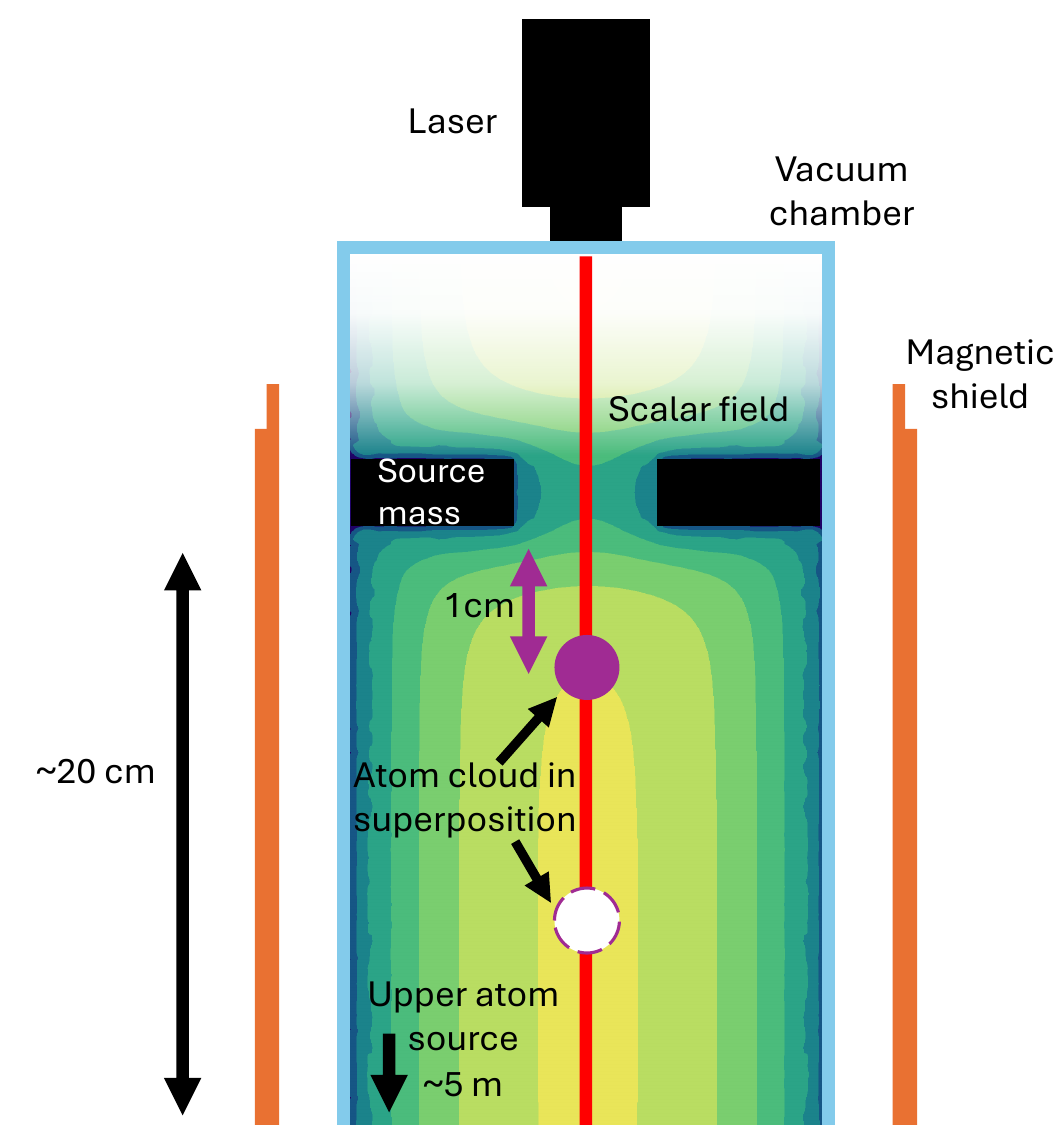}
    \caption{Schematic of the top of the proposed experimental setup (not to scale) showing the upper interferometer of the gradiometer. A $1\,\mathrm{cm}$ thick planar source mass is positioned at the top of the vacuum chamber with a central hole of radius 40\% of that of the chamber to allow for laser transmission.
    An atom cloud superposition is shown with the upper state having a closest approach to the plate of $1$\,cm. The scalar force drops off exponentially in strength with distance such that it barely acts on the lower state. These atoms are launched from approximately $5$\,m below the source mass. A second identical interferometer, with atom source at 0\,m, is operated simultaneously using common laser pulses (depicted by the red line). Magnetic shielding surrounds the vacuum chamber, protecting from stray fields.
    }
    \label{fig:experiment-diagram}
\end{figure}

A diagram of the proposed experimental set-up is shown in Fig.~\ref{fig:experiment-diagram}. The experiment comprises a 10.1\,m vertical baseline~($L)$ atom gradiometer consisting of two atom interferometers based on single-photon transitions in~$^{87}$Sr. An additional planar source mass is situated at the top of the vacuum chamber. In this gradiometer configuration, the upper and lower interferometers are run simultaneously using a common laser source.  Atom clouds are launched upwards in a fountain configuration from sources at 5\,m and 0\,m respectively, falling under gravity throughout the measurement sequence. We assume the atoms propagate up to 5\,m above each source, with the upper interferometer atoms reaching a closest approach of 1\,cm to the source mass. An additional 10\,cm of vacuum chamber accommodates the source mass and laser optics. The laser is delivered from the top of the chamber, with counter-propagating pulses realised via a retro-reflection mirror at the bottom of the vacuum chamber.

The key design parameters that we assume for the gradiometer are given in Table~\ref{tab:params1}, matching standards adopted by the TVLBAI community for forecasts~\cite{TVLBAI:2024, Abdalla:2024sst}. Whilst we remain agnostic to specific experimental designs, we anticipate that our conclusions will be applicable to 10\,m-scale gradiometers, such as AION-10~\cite{Bongs:2025rqe} or VLBAI~\cite{Hartwig:2015iza}, and could straightforwardly be adapted to longer baseline experiments, such as MAGIS-100~\cite{MAGIS-100:2021etm}.

{We assume a source mass thickness of $1\,\mathrm{cm}$, constructed of 316 stainless steel with density $\rho \approx 8\,\mathrm{g/cm}^3$. This material is selected for its much lower magnetic permeability than the more common 304 stainless steel~\cite{wilson1991magnetic}.\footnote{While existing gradiometer components, such as a retroreflector mirror or lens, could serve as the source mass, their material properties may be suboptimal. Moreover, if positioned outside of the magnetic shielding, atoms approaching closely enough to experience a significant fifth force may enter unshielded regions, potentially introducing phase noise from stray fields.} Since the atom clouds are addressed by counter-propagating laser pulses from opposing ends of the baseline, the source mass needs to allow for laser transmission. The simplest option is an annular plate with a concentric inner hole. In general, the smaller the hole the larger the gradient of the scalar field close to the plate, and in turn the larger the force  experienced by the atoms. This effect needs to be balanced against the onset of laser aberration and diffraction, which become more severe at small hole radii. A further consideration is beam clipping, which occurs if the beam size, which should be large enough to address the entire cloud of atoms exceeds the hole dimensions. It was argued in Ref.~\cite{Garber:2024strontium} that beam diffraction is significant when the hole radius is less than 40\% of the vacuum chamber radius. We therefore adopt $R_\mathrm{H} = 0.4 R_\mathrm{vac}$ as our fiducial standard for the remainder of this study, taking the inner radius of the vacuum chamber to be $R_\mathrm{vac} = 5$~cm.

\begin{table}[!t]
    \caption{Design parameters characteristic of a representative 10~m atom gradiometer experiment utilising $^{87}$Sr atoms. $k$ is the wavenumber of the laser; $L$ is the baseline; $n_{\rm max}$ is the maximum number of laser pulses; $\delta\Phi$ is the 
    phase noise.
    }
    \begin{tabular}{c c c c c} 
    \toprule
     Isotope \,& \,$k$ [m$^{-1}$]\, & \,$L$ [m] \,& \, $n_{\rm max}$  \,& \,$\delta\Phi$ [rad/$\sqrt{\mathrm{Hz}}$]  \\ 
    \midrule
     ${}^{87}$Sr & $9.002\times10^6$ & $10.1$ & $4000$ &$10^{-3}$  \\
    \bottomrule
    \end{tabular}
    \label{tab:params1}
\end{table}

\subsection{Numerical calculation of the scalar field profile}
\label{sec:numerics}

\begin{figure}[!t]
    \centering
    \includegraphics[width=0.7\columnwidth]{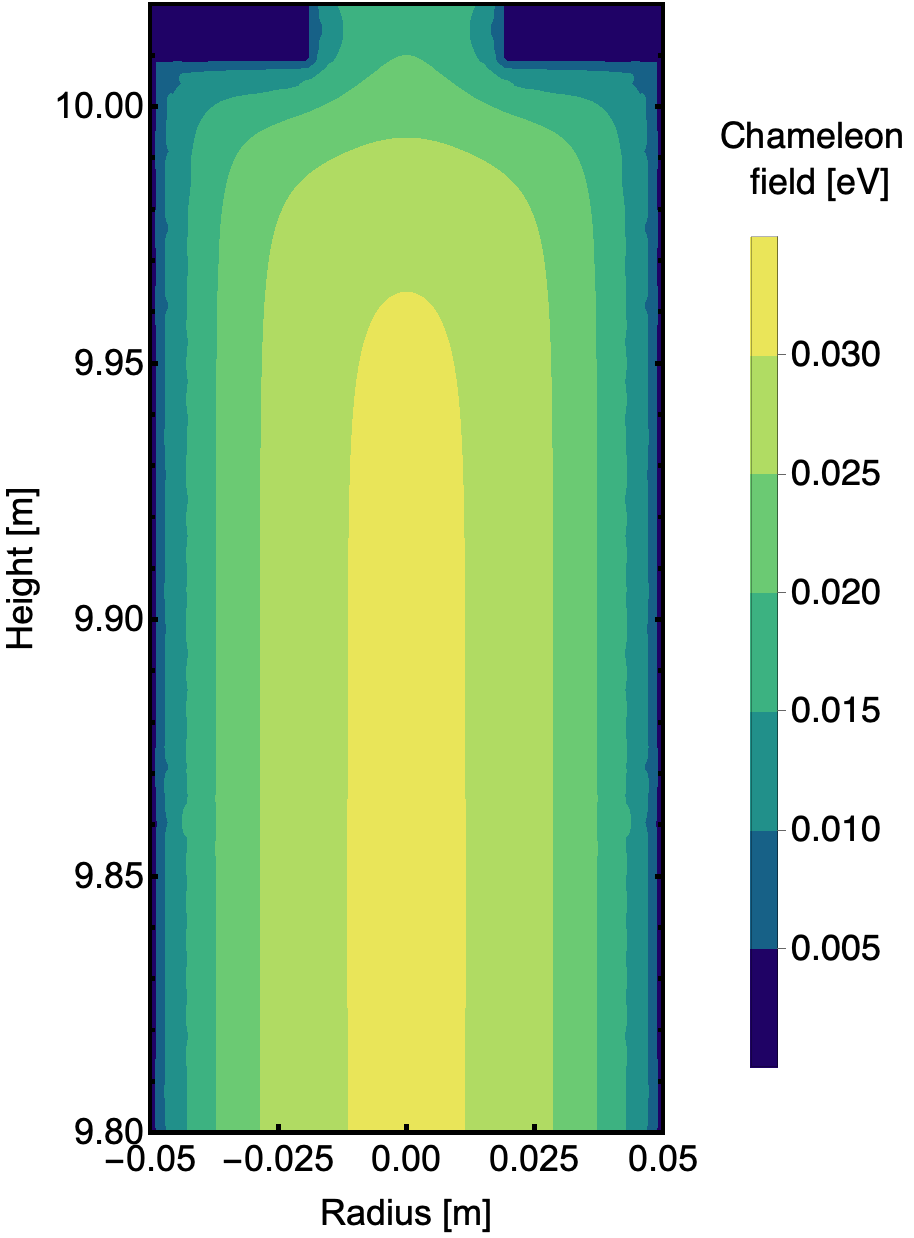}
    \includegraphics[width=0.7\columnwidth]{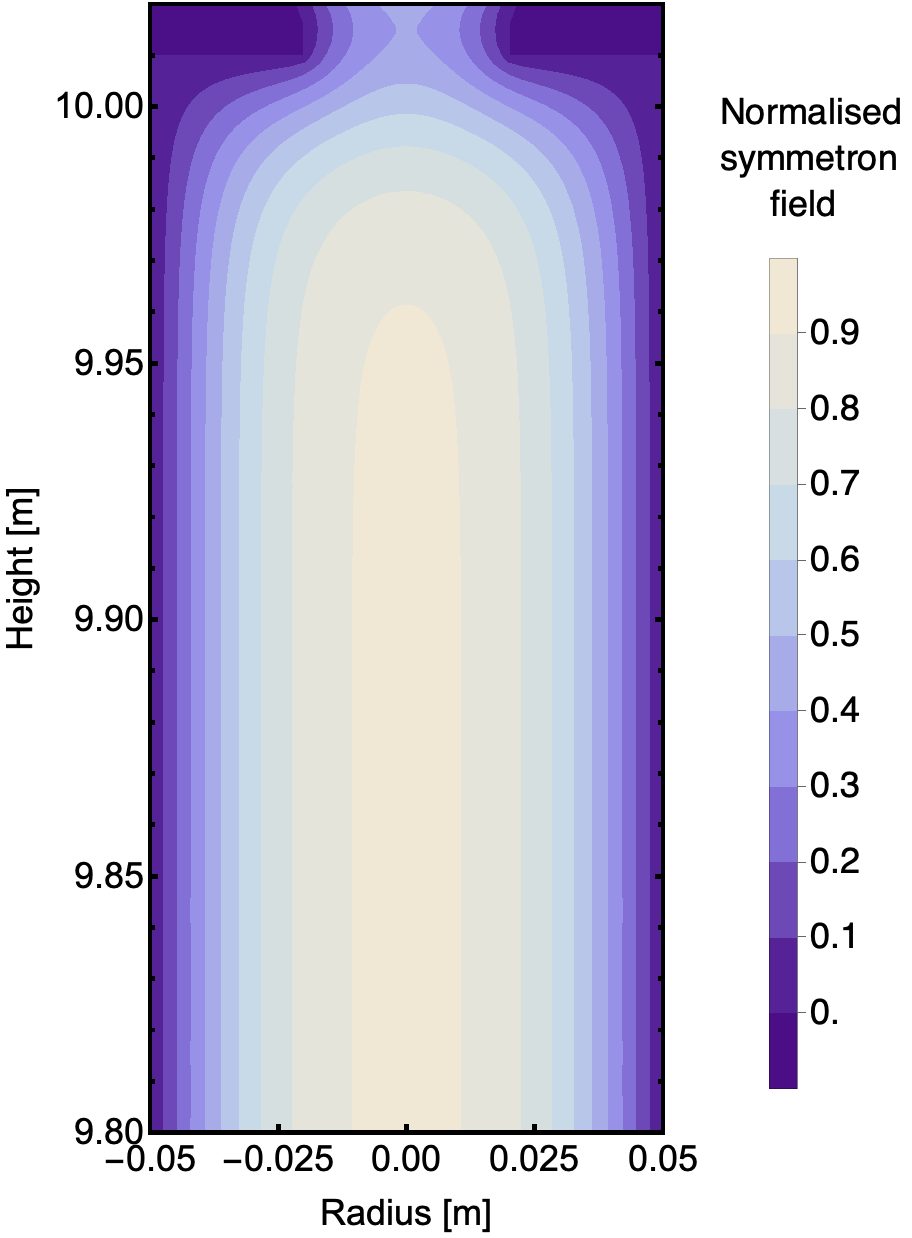}
    \caption{Numerically modelled scalar fields in a vacuum chamber, sourced by a $1$~cm thick plate at the top with a hole satisfying $R_{\mathrm{H}}=0.4 R_{\rm vac}$. {\bf Top}: Chameleon field with $M=M_{\rm Pl}$ and $\Lambda=10$\,meV. {\bf Bottom}: Symmetron field, normalised by its VEV~($v=\mu / \sqrt \lambda$), with $M=1$\,GeV, $\lambda=1$, and $\mu=1.5\times10^{-2}$\,meV.
    }
    \label{fig:fields}
\end{figure}

The fifth force experienced by the atoms at each point along their trajectories is governed by the local value of the scalar field. As such it is first necessary to compute the profile of the scalar field within the vacuum chamber. This amounts to solving the scalar field equation of motion (Eq.~\eqref{scalar-eom}) accounting for the residual gas density in the chamber, which we take to be $\rho_\mathrm{m}(\vec x) = 10^{-19}\,\mathrm{g/cm}^3$. The influence of the vacuum chamber walls and the annular source mass on the field is incorporated via boundary conditions imposed at their surfaces. 
The density of the atoms themselves is neglected in this computation, but later accounted for via the screening factor~$\lambda_a$. 
Once~$\varphi(\vec x)$ is determined, we use it to evaluate the scalar fifth force along the trajectory of the atoms via Eq.~\eqref{eq:acc}.

We solve the scalar field equation of motion using a finite element solver in Mathematica version 14.0.0. It is computationally advantageous to first non-dimensionalise the scalar equation of motion by making an appropriate transformation of variables.  For the chameleon this amounts to performing the rescalings
\begin{equation}
    \varphi \to \varphi / \Lambda \quad {\rm and} \quad \vec x \to \Lambda \vec x~,
\end{equation}
and for the symmetron
\begin{equation}
    \varphi \to \varphi / v \quad \mathrm{and} \quad \vec x \to \mu \vec x~.
\end{equation}
Given that the system is rotationally symmetrical about the polar axis, we work in cylindrical co-ordinates $(r,z)$ and solve in the region
\begin{align} \nonumber
    0 < r < R_\mathrm{vac}~, \\
    z_\mathrm{min} < z < z_\mathrm{max}~.
\end{align}

The scalar force is exponentially suppressed beyond distances $R_\mathrm{vac}$ from the plate, so it is not necessary to simulate the entire length of the vacuum chamber.  This suppression occurs for slightly different reasons in the two theories.  For chameleon theories, this arises because chameleon perturbations have an `effective mass' $m_\mathrm{eff}$ (see Appendix~\ref{app:boundary-conditions}) inside the vacuum chamber that is at most $m_\mathrm{eff} \approx R_\mathrm{vac}^{-1}$.
For the symmetron, the theory contains a mass parameter $\mu$.  If $\mu < R_\mathrm{vac}^{-1}$ then the ambient symmetron field is driven to zero inside the vacuum chamber~\cite{Upadhye:2012rc} such that the fifth force vanishes.  In both cases, the scalar field acquires an effective mass such that the Compton wavelength of the field, given by the inverse mass, is approximately~$R_\mathrm{vac}$. This results in an exponential suppression of the force beyond this distance. The vacuum chamber radius thus sets an upper limit on the interaction distance between the atoms and the source mass, a fact which will prove of use in calibrating the experiment.

As a result of the exponential suppression of the force beyond distances $R_\mathrm{vac}$ from the plate, we only simulate the top $21.5\,\mathrm{cm}$ of the vacuum chamber.  The top of the simulation box corresponds to the centre of the source mass, i.e., $z_\mathrm{max} = 10.015\,\mathrm{m}$.  The lower limit is $z_\mathrm{min} = 9.8\,\mathrm{m}$.

Having defined the computational domain, we now specify the boundary conditions. 
We impose $d \varphi/dr = 0$ at $r = 0$  and $d \varphi / dz = 0$ at $z = z_\mathrm{min}$, as well as at $z = z_\mathrm{max}$ when $r \leq R_\mathrm{hole}$.  We set $\varphi = 0$ at $r = R_\mathrm{vac}$ and on the surface of the source mass. These boundary conditions are justified in Appendix~\ref{app:boundary-conditions}.  As argued there, the condition $\varphi = 0$ on the surface of the vacuum chamber walls and source mass is an approximation that holds well within the theory parameter spaces of interest.  Were this condition not satisfied, macroscopic objects would be unscreened and such models would already be subject to stringent constraints from both laboratory experiments and astrophysics on Yukawa-like fifth forces.

Figure~\ref{fig:fields} shows characteristic numerical solutions of the chameleon and symmetron. A separate numerical solution must be obtained for each ($\Lambda, M$) of interest for the chameleon and each ($\mu, M$) for the symmetron. The dimensionless coupling $\lambda$ in the symmetron theory scales out of the numerical calculation of the field but re-enters later when computing the acceleration of the atoms.\footnote{For the symmetron, the ambient gas density inside the vacuum chamber was neglected. 
This has negligible effect provided the gas density is less than the symmetron critical density $\rho_\mathrm{crit}$. If $\rho_\mathrm{gas} > \rho_\mathrm{crit}$ the scalar field remains at $\varphi = 0$ everywhere inside the vacuum chamber and the force is zero.  To account for this, we set $\phi = 0$ by hand whenever $\rho_\mathrm{gas} > \rho_\mathrm{crit}$, which simplifies the numerical calculations.}

The qualitative features of Fig.~\ref{fig:fields} may be understood in the following way. In both the chameleon and symmetron models, the scalar field $\varphi$ is suppressed in dense environments and takes larger values in under-dense regions. In our setup, the dense source mass and vacuum chamber walls therefore reduce $\varphi$ near their surfaces. As shown in Fig.~\ref{fig:fields}, the field is small near the source mass and at the chamber edges, and rises towards a constant value at the centre of the vacuum chamber. This spatial variation generates a gradient in the scalar field that, via Eq.~\eqref{eq:acc}, produces an attractive force between the source mass and the atoms. For the models of interest this force either scales as $\vec \nabla \varphi$ or $\varphi \vec \nabla \varphi$.
As a result, the regions in Fig.~\ref{fig:fields} where the field value is greatest do not necessarily correspond to the locations of strongest force. 

With the solutions for the scalar field in hand, the integrated phase accrued along the trajectory of the atoms, and in turn the gradiometer phase, can be computed, as we now detail.

\subsection{Gradiometer phase}
\label{sec:ep}

The observable of an atom gradiometer experiment is the difference in the phase differences recorded by the upper ($u$) and lower ($l$) interferometers at the end of the interferometry sequence:
\begin{equation}
\label{eq:grad}
\Delta \Phi = \Phi^u - \Phi^l~.
\end{equation}
The gradiometer phase is sensitive to the relative acceleration between the two atom clouds. Since the interferometers operate at different heights, the upper and lower interferometers experience differing gravitational and scalar interactions with the plate. 

The trajectory of atoms propagating under the Earth's gravitational field is described by
\begin{equation}
    L = m_a\left(\frac{1}{2}\dot{z}^2-gz+\frac{1}{2}\gamma_{zz}z^2\right)~,
    \label{eq:traj}
\end{equation}
where $m_a$ is the mass of the atoms, $g$ is the local acceleration due to the gravity of the Earth, and $\gamma_{zz}$ is the Earth's vertical gravity gradient.
We neglect effects from Coriolis forces and assume they are corrected with a rotation compensation system (see, e.g., Ref.~\cite{MAGIS-100:2021etm, Bongs:2025rqe}).

To compute the contribution of the screened scalar field to the phase, we use the semi-classical perturbative approach~\cite{Chu:2001gct, Bongs:2002, Overstreet:2020ftt, Storey:1994oka}, which gives
\begin{equation}
    \Phi^{(u,l)} = \frac{m_a}{\hbar}\int_0^{2T}{\rm d}t\,\left[V_{p,\varphi}\left(z^{(u,l)}_2\right)-V_{p,\varphi}\left(z_1^{(u,l)}\right)\right]~.
    \label{eq:phase}
\end{equation}
Here, $T$ is the interrogation time while $z_1 (t)$ and  $z_2 (t)$ are the trajectories of the two interferometer arms, unperturbed by potentials sourced by the plate, following from  Eq.~\eqref{eq:traj} and the laser pulse sequence. $V_{p,\varphi}$ is the potential energy per unit mass experienced by the atoms from the scalar field. Since the atoms propagate in the $z$ direction, this is calculated by
integrating the $z$-component of Eq.~\eqref{eq:acc}, which gives
 \begin{equation}
    V_{p,\varphi}(z) = \int_0^z{\rm d}z^\prime\, \lambda_a \frac{\rm d}{{\rm d} z^\prime} A(\varphi)~.
    \label{eq:potential_int}
\end{equation} 
The scalar field contribution to the gradiometer phase, $\Delta \Phi_{\varphi}$, then follows from Eq.~\eqref{eq:grad}. 

The gradiometer phase resulting from the plate's gravitational field, $\Delta \Phi_g$, is obtained analogously by replacing $V_{p,\varphi}$ in Eq.~\eqref{eq:phase} with the Newtonian potential along the trajectory of the atoms (the central axis of the annular plate): 
\begin{equation}
\begin{split}
    V_{p,g}(z) = -2\pi G_N \rho  W \bigg(&\sqrt{[z(t)-h]^2+R_\mathrm{vac}^2}\\
    &-\sqrt{[z(t)-h]^2+R_{\rm{H}}^2}\bigg)~.
\end{split}
\label{eq:ring_potential}
\end{equation}
{Here, $R_\mathrm{vac}$ is the radius of the plate, $R_{\mathrm{H}} = 0.4 R_\mathrm{vac}$ is the radius of the hole, $z(t) - h$ is the distance between the atoms and the centre of the plate,~$\rho$ is the density of the plate, and~$W$ is the plate thickness.}
For the parameter regions of interest, $\Delta \Phi_g$ dominates over~$\Delta \Phi_{\varphi}$.

\begin{figure*}[!t]
    \centering
    \includegraphics[width=0.96\columnwidth]{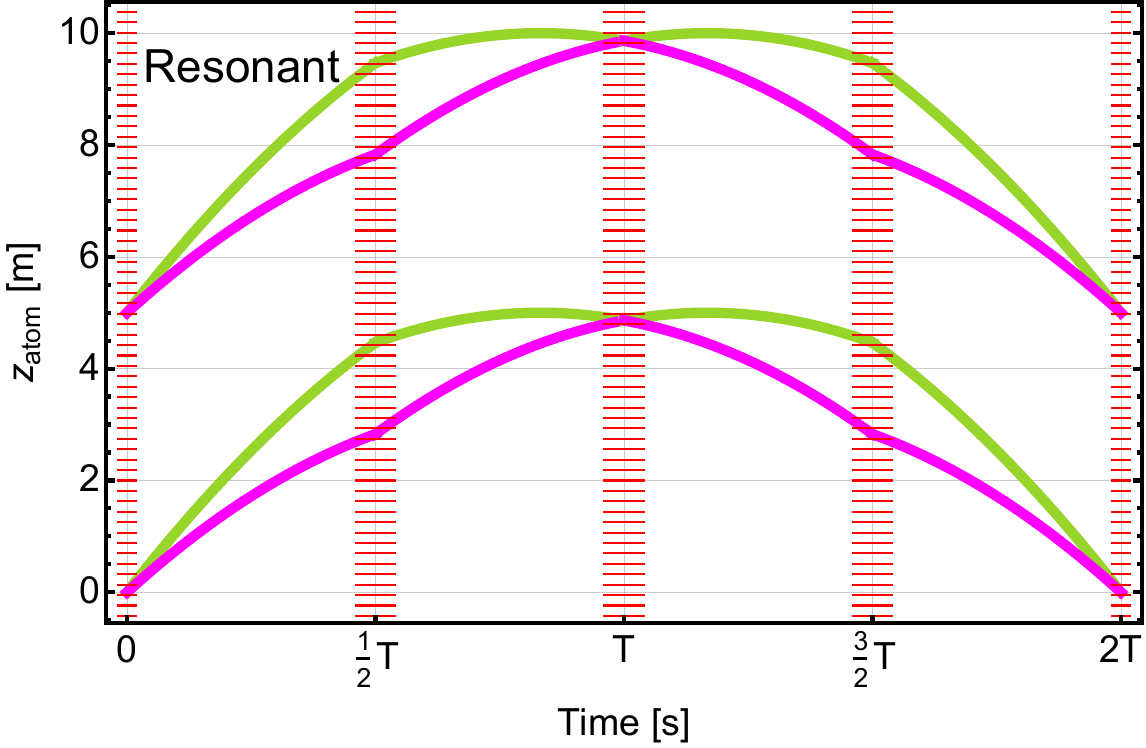}
    \includegraphics[width=1.01\columnwidth]{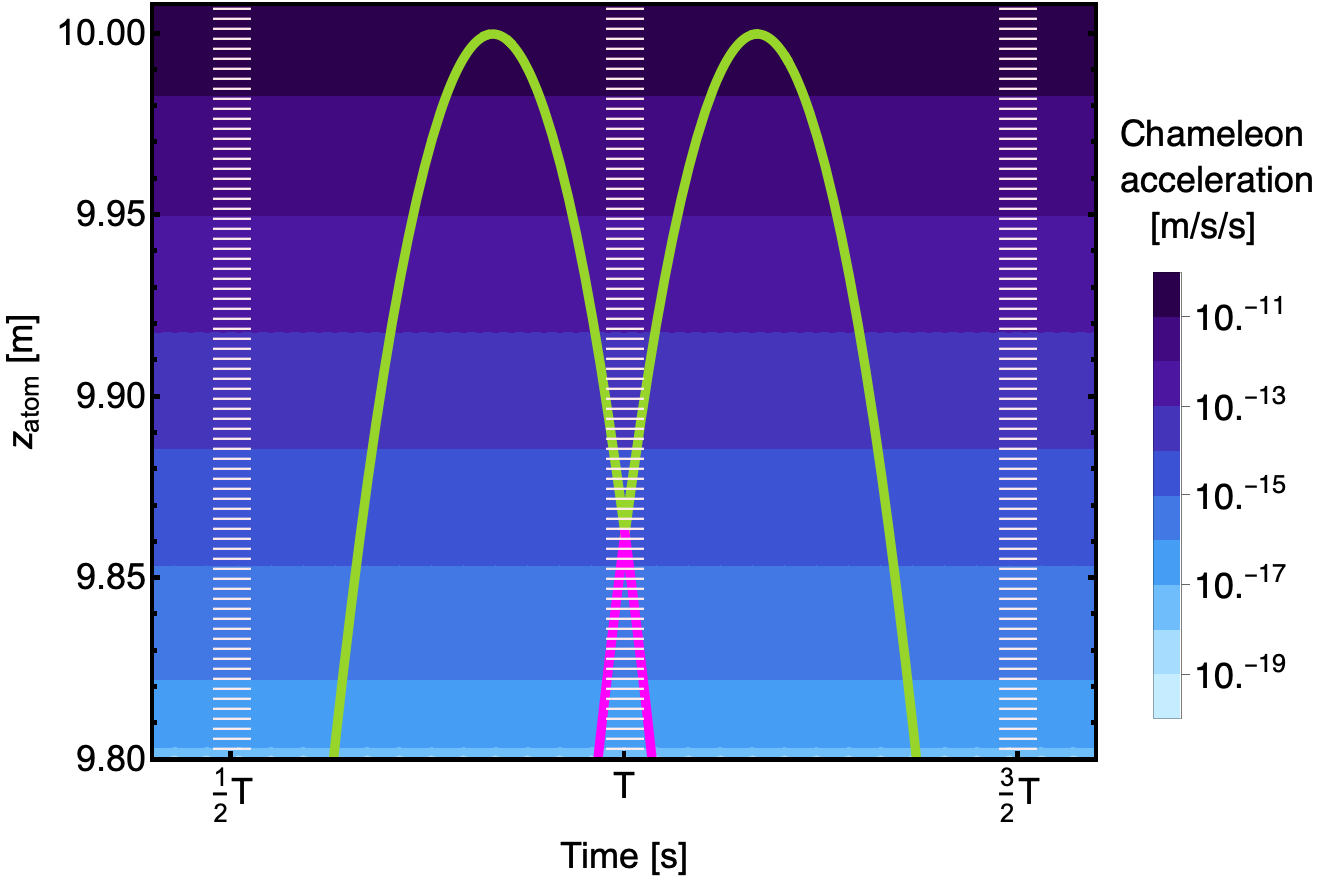}
    \includegraphics[width=0.96\columnwidth]{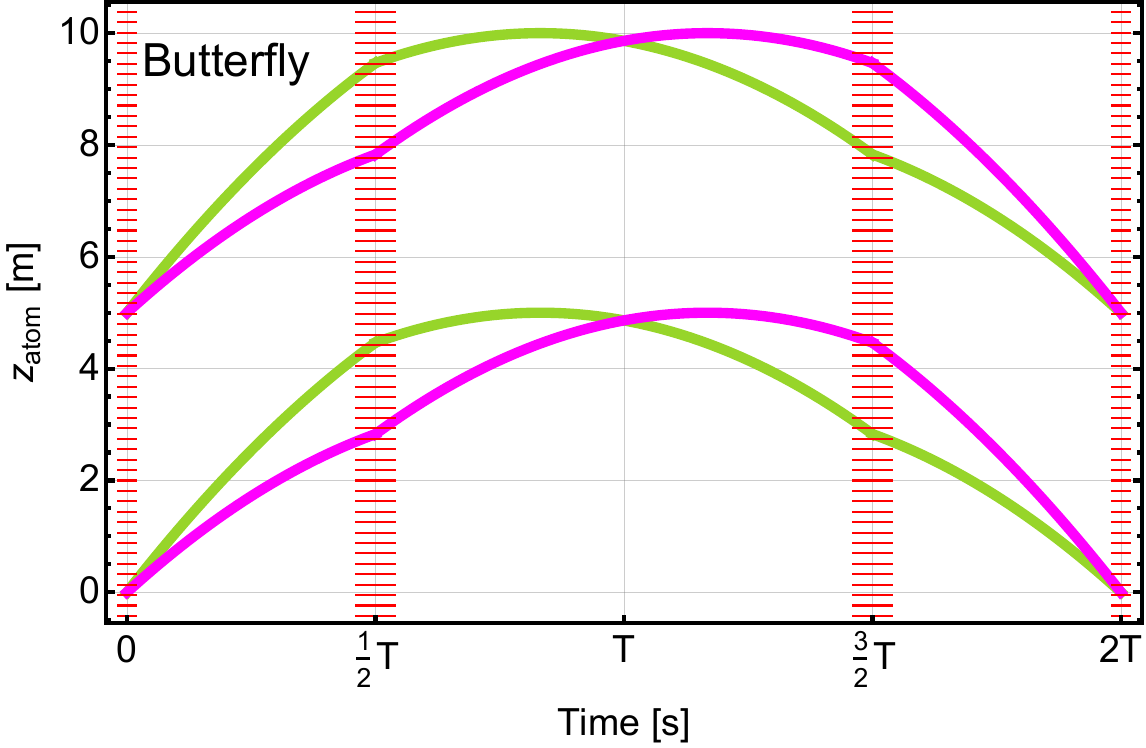}
    \includegraphics[width=1.01\columnwidth]{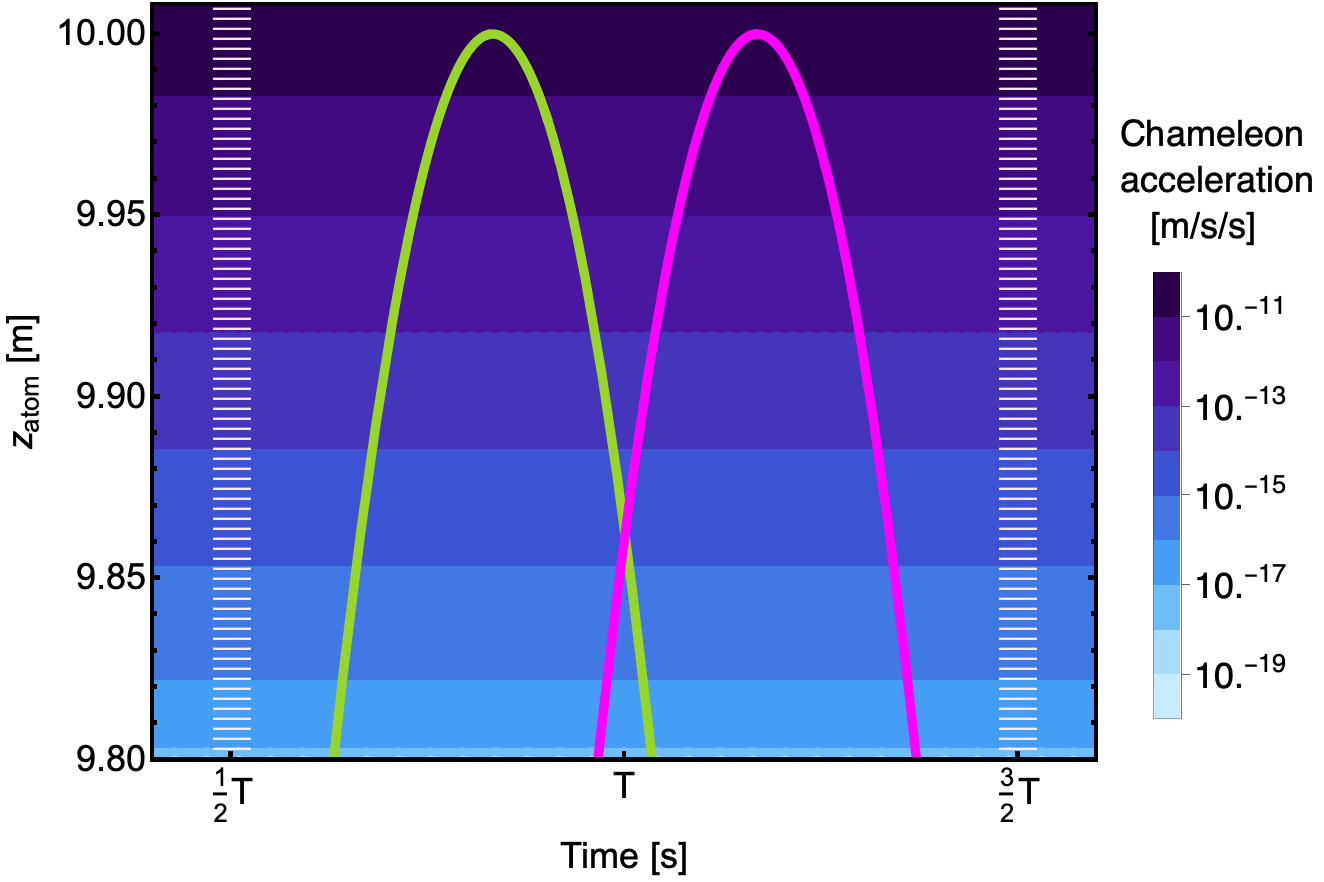}
    \caption{Spacetime diagrams of atom gradiometer sequences in the presence of a plate at $10.01$~m. The acceleration sourced by a chameleon field with $M=M_{\rm Pl}$ and $\Lambda=\mathrm{meV}$ is shown by the blue contours in the right panels.
    {\bf Top}: Spacetime diagram of a resonant $Q=2$ sequence. 
    The left panel shows the full sequence with pulses at times $0$,  $T/2$, $T$, $3T/2$, and $2T$. The right panel shows the atom trajectories in the vicinity of the plate.  {\bf Bottom}: Spacetime diagram for the butterfly sequence, which is insensitive to scalar forces due to the symmetry of the sequence. The right panel shows the atom trajectories in the vicinity of the plate. The resonant and butter sequences differ by the omission of the central laser pulses at time $T$.
    }
    \label{fig:Q2}
\end{figure*}

\subsection{$Q$-flip protocol}
\label{sec:Qflip}

Given the falloff of the scalar field gradient with distance from the plate, the resulting energy potential per unit mass (Eq.~\eqref{eq:potential_int}) is strongly localised near the source mass. 
To maximise the accumulated phase (Eq.~\eqref{eq:phase}), the atom trajectories (of the upper interferometer) should therefore remain close to the plate for as long as possible. Sufficient spatial separation between the two interferometer arms should also be maintained such that only the upper arm experiences a significant force from the scalar field.

The trajectories of the atomic wave-packets are determined by the sequence of applied laser pulses.  Whilst typical broadband large-momentum transfer (LMT) Mach-Zehnder sequences
provide excellent separation between the interferometer arms~\cite{Hogan:2008}, their central mirror pulse accelerates the upper arm away from the source mass, reducing the time spent close to the plate. Instead we employ a resonant sequence with multiple closed spacetime diamonds~\cite{Graham:2016plp}, where atoms receive additional momentum kicks towards the source mass. In such sequences, the number of closed spacetime diamonds is denoted by $Q$ and the total number of laser pulses is 
$ 2 Q (2n-1)+1$,
consisting of two $\pi/2$-pulses that define the start and end of the sequence, and $[2Q(2n-1)-1]$ $\pi$-pulses. 
The number of LMT-kicks during a single diamond in the sequence is denoted by~$n$.

The top row of Fig.~\ref{fig:Q2} shows a resonant sequence with $Q=2$, in which the atom trajectories in the upper interferometer peak at approximately $10$\,m before and after the central mirror pulse at time $T$. The additional momentum kicks in the $Q=2$ sequence increase the overall spacetime area in the presence of the scalar force compared to a Mach-Zehnder sequence. The lower arm is also kicked closer to the source mass; however, as seen in the top right panel, the greatest force to which the lower arm is subject is several orders of magnitude smaller than the maximum force experienced by the upper arm.
Whilst $Q>2$ resonant sequences increase the time that the atoms spend close to the plate, the separation between the upper and lower interferometer arms is reduced.  We find that these sequences provide no advantage and therefore adopt the $Q=2$ sequence. To maximise the total number of pulses within the design constraint $n_{\rm max}=4000$, we set $n=500$.

\begin{figure}[!t]
    \centering
    \includegraphics[width=\linewidth]{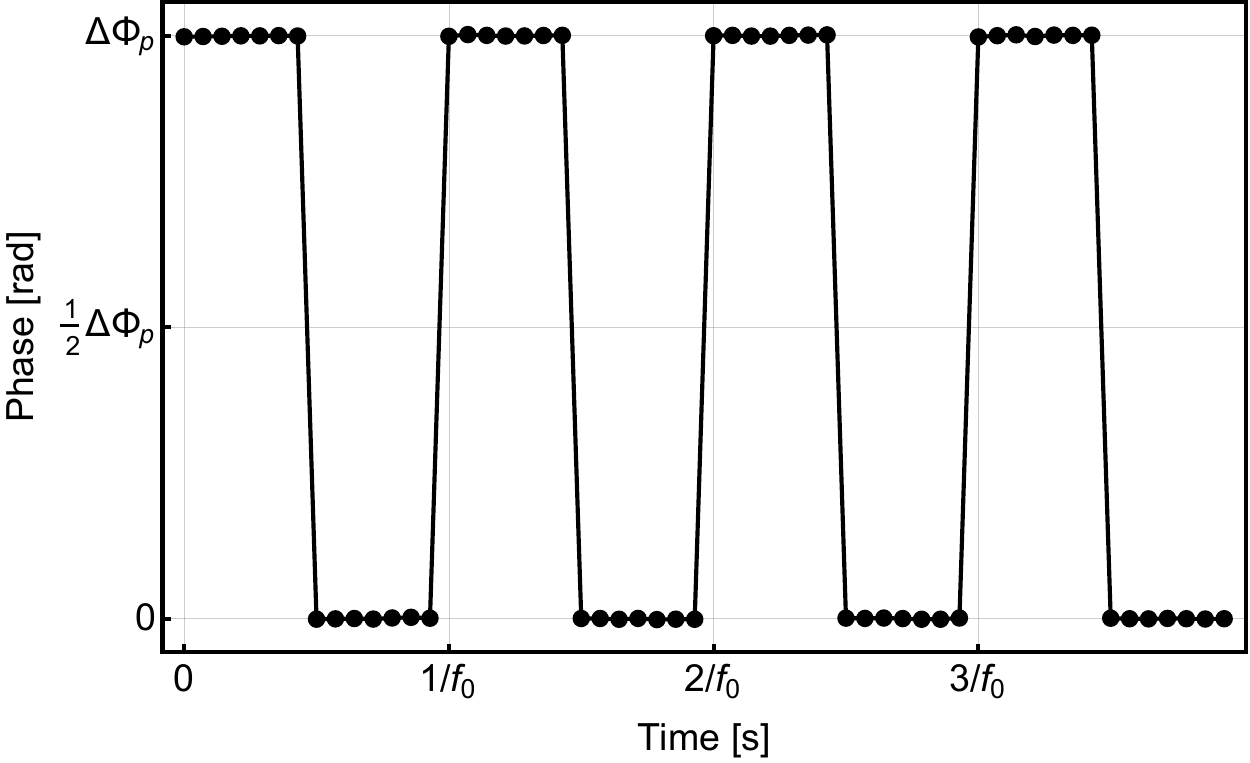}
    \includegraphics[width=\linewidth]{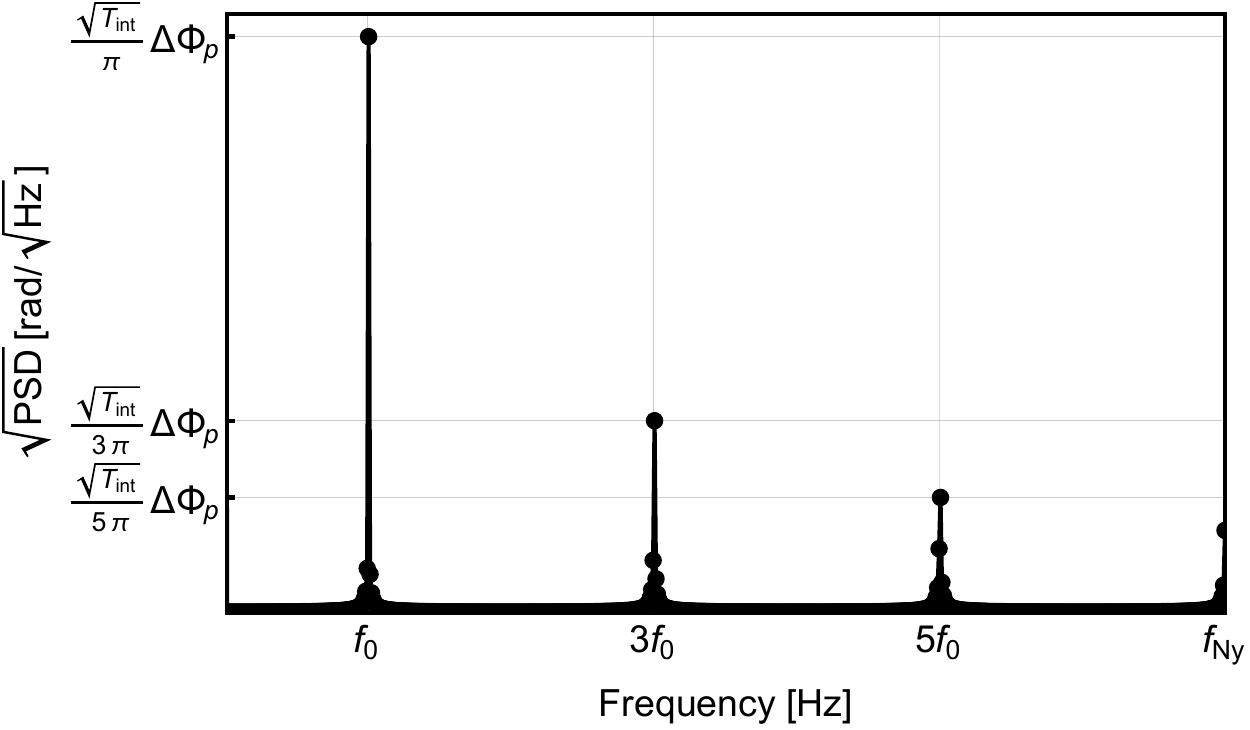}
    \caption{{\bf Top}:  Time series of gradiometer phases using the $Q$-flip protocol. Measurements are first made using the resonant $Q=2$ sequence, which is sensitive to the scalar and gravitational fields sourced by the plate. The sequence then switches to the butterfly sequence, which is insensitive to these effects. Stochastic shot noise is present in each measurement. Repeating the resonant--butterfly cycle produces an oscillating phase readout with period $1/f_0$. {\bf Bottom}: Square-root of the PSD of the above time series. Spikes appear at odd multiples of the fundamental frequency $f_0$ with amplitudes proportional to  $\sqrt{T_{\rm int}}\Delta\Phi_p$.}
    \label{fig:oscillating_signal}
\end{figure}

In fifth force experiments, systematic effects that are not associated with the source mass are typically removed by calibrating with measurements made in the absence of the source mass. To circumnavigate the engineering challenges associated with removing and replacing the source mass from the vacuum chamber and the additional systematics this re-positioning is likely to introduce, we instead achieve this calibration using a `butterfly' sequence~\cite{McGuirk:2002zz, Canuel:2006zz}, which is insensitive to the presence of the plate. This sequence is shown in the lower panels of Fig.~\ref{fig:Q2}. The butterfly sequence uses the same value of $n=500$ as the $Q=2$ resonant sequence but omits the laser pulses at time $T$. This causes the interferometer arms to cross over, with the result that each arm spends the same amount of time near the plate.

Since the source mass is fixed in place, the fifth force is constant in time. Such a static effect is more difficult to detect than inherently oscillatory phenomena, such as GWs or ULDM, whose characteristic time dependence can be used to distinguish them from static or slowly varying backgrounds. To overcome this challenge, we induce a controlled time dependence in the fifth-force signal by alternating between two pulse sequences during a continuous experimental run: the resonant $Q=2$ sequence, which is sensitive to scalar fields sourced by the plate, and the butterfly sequence, which is not. Periodically switching between them generates an oscillatory phase response at frequency $\omega_0 = 2\pi f_0$, which can be controlled experimentally. We coin this the `$Q$-flip protocol'.

This effect is illustrated schematically in the time series shown in the top panel of Fig.~\ref{fig:oscillating_signal}, which includes contributions from the plate’s gravitational field, a screened scalar field, and stochastic shot noise. The resonant-sequence has a mean amplitude $\Delta\Phi_p = \Delta\Phi_g + \Delta\Phi_\varphi$, around which stochastic shot-noise fluctuations occur. In this schematic, the shot-noise fluctuations are assumed to be much smaller than~$\Delta\Phi_p$.
Since the gravitational and scalar contributions are absent in the butterfly sequence, its output fluctuates around zero due to atom shot noise.

The periodicity of the signal concentrates into peaks in frequency space~\cite{schelfhout2021fourier}. 
This allows transient noise sources that remain broadly distributed to be effectively de-trended using standard techniques for detecting oscillatory signals. This behaviour is formally captured by the power spectral density (PSD)  of the phase measurements. For a set of~$N$ discrete measurements $\{\Delta \Phi_\ell\}$ recorded at times $t = \ell \Delta t$, where $\Delta t$ is the time separation between successive measurements, over a total integration time $T_{\rm int} = N \Delta t$, the PSD is defined~as
\begin{equation}
    S_k  = \frac{\Delta t}{N}\left|\sum_{\ell=0}^{N-1} \Delta\Phi_\ell \exp\left[-\frac{2 \pi i \ell k}{N}\right]\right|^2\;,
    \label{eq:PSD}
\end{equation}
where $k=0,\dots,(N-1)$ labels the discrete frequencies $f_k = k /T_\mathrm{int}$ with resolution $\Delta f = (T_\mathrm{int})^{-1}$. The maximum non-aliased frequency is known as the Nyquist frequency, $f_\mathrm{Ny} = (2\Delta t)^{-1}$.

The lower panel of Fig.~\ref{fig:oscillating_signal} shows the square root of the PSD (the amplitude spectral density) for the upper-panel time-series. The fundamental peak occurs at $f_0$, followed by smaller spikes at odd multiples of $f_0$. While these higher harmonics are not exploited in the present analysis, they could provide additional information for signal characterisation in the event of a detection.
In practice, the peaks have finite widths, but these become negligible provided sufficiently long integration times, $T_{\rm int} \gg 1/f_0$, and their amplitudes in this limit scale as $\sqrt{T_{\rm int}}$. In the following, we focus on the spike at the fundamental frequency.

To evaluate the sensitivity to the screened scalar models, we define the signal-to-noise ratio (SNR) as the ratio of the PSD of the scalar field contribution, $S^{\varphi}$, to the PSD of the background noise, $S^n$, evaluated at the fundamental frequency:
\begin{equation}
    {\rm SNR} = \frac{S^{\varphi}(\omega_0)}{S^n(\omega_0)}\,.
    \label{eq:SNR2}
\end{equation}
Note that we have implicitly rewritten the PSDs as a (continuous) function of $\omega = 2\pi f$, which is a valid approximation provided $T_{\rm int} \gg 1/f_0$.

In the following, we assume a periodic signal with amplitude $\Delta\Phi_{\varphi}$ in the limit $T_{\rm int} \gg 1/f_0$, which yields $
 S^{\varphi}(\omega_0)\approx \pi ^{-2} T_{\rm int}|\Delta\Phi_\varphi|^2$. 
We assume that the background noise is dominated by atom shot noise, which is white in frequency space, with $S^n(\omega_0)=|\delta\Phi|^2$, where $\delta\Phi$ is a design parameter quoted in Table~\ref{tab:params1}. In a practical experiment, $S^{\varphi}(\omega_0)$ must be extracted from the square-root of the measured PSD by removing the contribution from the plate's Newtonian gravity.  We next discuss a protocol for doing this.

\subsection{Calibration and source mass calibration}\label{sec:calibration}

The $Q$-flip protocol measures the difference in gradiometer phases between resonant and butterfly sequences. This is sensitive to the \textit{total} potential sourced by the plate, with Newtonian gravity providing the dominant contribution. The peak at the fundamental frequency in the $\sqrt{{\rm PSD}}$ has amplitude $ \pi^{-1}\sqrt{T_{\rm int}}\Delta \Phi_p = \pi^{-1}\sqrt{T_{\rm int}}  (\Delta \Phi_g + \Delta \Phi_\varphi )$.  Isolating~$\Delta \Phi_\varphi $ requires accurate subtraction of $\Delta \Phi_g$. Uncertainty in this subtraction, $\delta[\Delta \Phi_g]$, directly limits the sensitivity to screened scalars. To achieve a sensitivity that is limited by shot noise, we require $ \pi^{-1}\sqrt{T_{\rm int}}  \delta[\Delta \Phi_g]\ll\delta\Phi$, where $\delta\Phi$ is the phase noise from Table~\ref{tab:params1}.

In principle, the~$\Delta \Phi_g$ contribution can be computed from Eqs.~\eqref{eq:phase} and~\eqref{eq:ring_potential} given the plate parameters $\{W,\rho,R_{\rm{vac}},R_{\rm{H}} \}$ and $G_N$. These could be determined from independent measurements with their uncertainties then propagating directly to the inferred value of~$\delta[\Delta \Phi_g]$. The dominant uncertainty arises from the plate density. The gravitational constant is known to a relative uncertainty $\delta G_N/G_N \approx 2.2 \times 10^{-5}$~\cite{Mohr:2024kco}, whilst we anticipate that the plate thickness and radius can be measured to $\delta W/W \sim \delta R/R \sim 10^{-4}$, corresponding to micrometre precision~\cite{salsbury2022test}. For 316 stainless steel, manufacturer specifications do not provide a tightly defined tolerance, so we adopt a conservative  relative uncertainty of $\delta \rho/\rho \sim 10^{-3}$, broadly consistent with the precision achieved in hydrostatic weighing of stainless-steel standards~\cite{HAYU2019120}. Under these conditions, $\delta[\Delta \Phi_g]\approx (\delta \rho/\rho) \Delta \Phi_g$. For $d_c=1\,\mathrm{cm}$,  where $d_c$ is the distance of the closest approach of the atoms to the plate, we find $\Delta \Phi_g \simeq 0.54\,\mathrm{rad}$ and thus
$\delta[\Delta \Phi_g] \simeq 5.4 \times 10^{-4}\,\rm{rad}$.
From the requirement $ \pi^{-1}\sqrt{T_{\rm int}}  \delta[\Delta \Phi_g]<\delta\Phi$, we find that for integration times exceeding $T_{\rm{int}}\simeq 34\,\mathrm{s}$, this method results in the sensitivity projections becoming limited by the uncertainty in the plate's gravitational contribution rather than atom shot noise.
 
Long-baseline gradiometers offer an approach to overcome this limitation and achieve improved sensitivity.  By controlling the atom launch velocity,~$d_c$ can be varied, allowing forces to be probed over a wide range of separations.  As illustrated in Fig.~\ref{fig:plate_characterise}, the scalar force drops off exponentially beyond its characteristic range ($\sim R_{\rm{vac}}$ as discussed in Sec.~\ref{sec:numerics}), whilst the gravitational force follows an inverse square law. Calibration measurements with the $Q$-flip protocol at large~$d_c$ values can therefore distinguish these contributions and characterise the plate's gravitational field \emph{in situ}.

To quantify the required calibration effort, consider performing measurements at $N_{\rm{cal}}$ different values of $d_c\gtrsim R_{\rm{vac}}$. At each value of $d_c$, the $Q$-flip protocol is run for a total integration time $T_{\rm{int}}$, giving a measurement of $\Delta \Phi_g$ with shot-noise-limited uncertainty $\delta[\Delta \Phi_g] = \pi \,\delta\Phi/\sqrt{T_{\rm{int}}} $. 
For the purposes of this estimate, we make the simplifying assumption that the dominant uncertainty on the gravitational phase follows from the product of terms $G_N \rho W$.
The $N_{\rm{cal}}$ measurements of $\Delta\Phi_g$ at different separations can therefore be fitted to the phase predicted by the gravitational model from Eq.~\eqref{eq:ring_potential} to constrain this overall scaling pre-factor. 
 Fitting to $N_{\rm{cal}}$ independent measurements at different $d_c$ values reduces the uncertainty on the fitted pre-factor by $1/\sqrt{N_{\rm{cal}}}$ compared to a single measurement, and thus the uncertainty on the predicted gravitational phase at $d_c=1\,\rm{cm}$ is $\delta[\Delta \Phi_g]/\sqrt{N_{\rm{cal}}}$. 
 To achieve shot-noise-limited sensitivity for the screened scalar search at the \emph{same integration time} as each calibration measurement requires $\delta[\Delta \Phi_g]/\sqrt{N_{\rm{cal}}} \ll \pi \,\delta\Phi/\sqrt{T_{\rm{int}}} $, or equivalently, $\sqrt{N_{\rm{cal}}}\gg 1$.

For $N_{\rm cal}=10$, performing measurements each integrated for $T_{\rm{int}}=12\,\rm{hours}$ yields a predicted gravitational phase at $d_c=1\,\rm{cm}$ with uncertainty $\delta[\Delta \Phi_g] \simeq 1.5 \times 10^{-5}\,\rm{rad}$. This is an improvement over what could be achieved through external measurements alone. This demonstrates that the \emph{in situ} calibration approach can achieve shot-noise-limited sensitivity even without exploiting prior information from laboratory measurements of the plate parameters. In practice, a Bayesian analysis incorporating priors on the geometric parameters $\{W,R_{\rm{vac}},R_{\rm{H}}\}$ and $\{\rho,G_N\}$ would yield tighter constraints and should be adopted to optimise the calibration efficiency.

\begin{figure}[!t]
    \centering
    \includegraphics[width=\columnwidth]{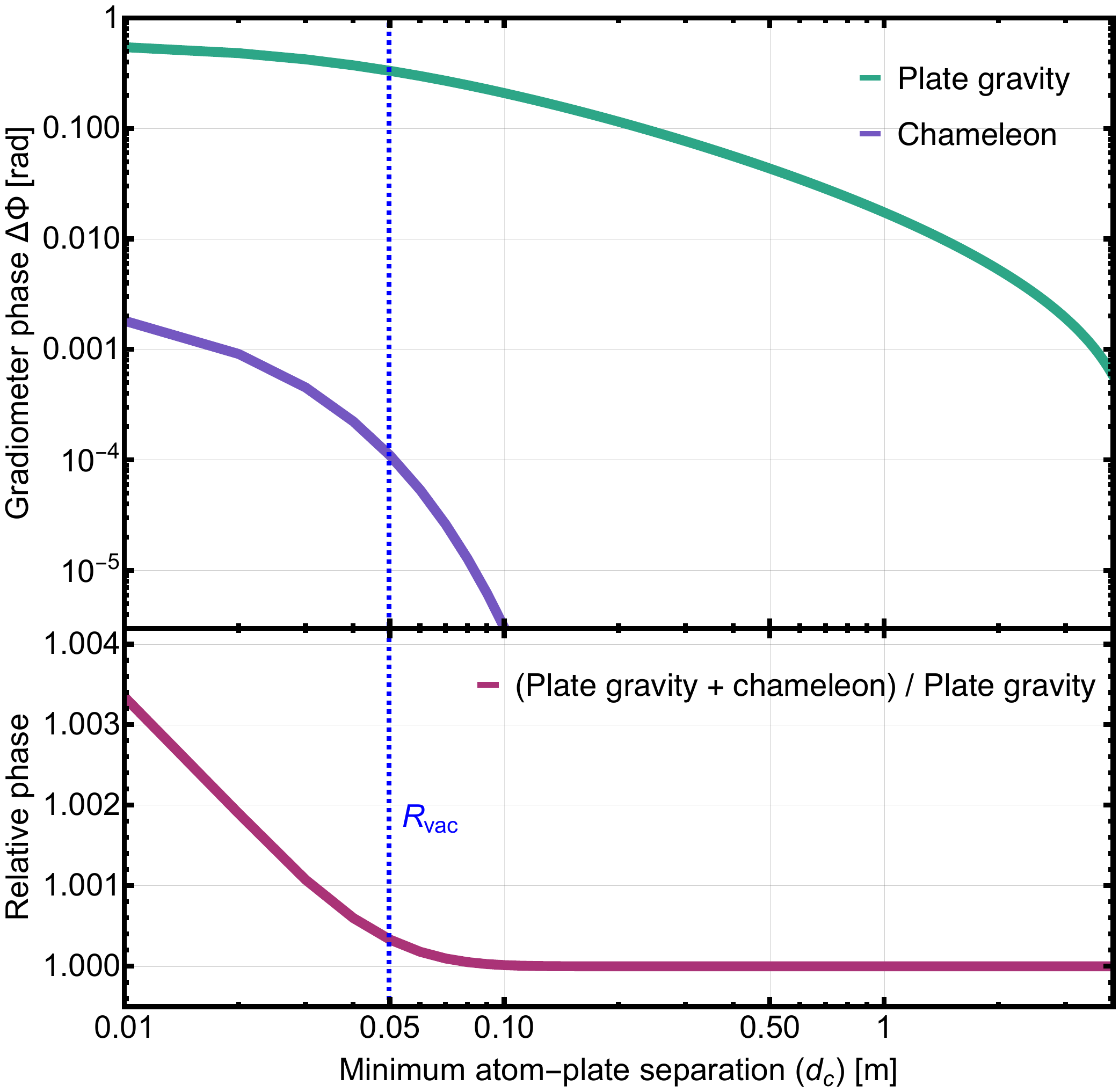}
    \caption{
     Gradiometer phase as a function of minimum atom-plate separation $d_c$, achieved by varying the atom launch velocity. Upper panel: plate gravity (green) and chameleon (purple) with $\Lambda=10^{-1.7}$\,meV, $M=10^{-4}\,M_\mathrm{Pl}$ contributions to the phase. Lower panel: relative phase with respect to the gravitational contribution. The blue dashed line shows $d_c = R_{\rm vac}$. Gravitational effects dominate the total phase, with the scalar force detectable only at $d_c \lesssim R_{\rm vac}$ due to exponential suppression at larger separations. By measuring the phase at multiple separations $d_c > R_{\rm vac}$, the gravitational contribution can be characterised and subtracted from measurements at $d_c = 1\,{\rm cm}$.
     }
    \label{fig:plate_characterise}
\end{figure}

\subsection{Plate hole effects}
\label{sec:plate-hole}

Our experimental proposal employs a static annular source mass with a central hole of radius $R_{\rm{H}}=0.4 R_{\rm{vac}}$ to allow for laser transmission. This constraint reduces the scalar field gradient near the source mass, and in turn the force on the atoms. A hole  smaller than $40\%$ of the vacuum radius, however, results in larger diffraction and aberration of the laser beam, introducing additional systematic effects~\cite{Garber:2024strontium}. Before proceeding to sensitivity projections, we quantify the impact of the hole geometry on the gradiometer phase for the screened models of interest.

\begin{figure}[!t]
    \centering
    \includegraphics[width=\columnwidth]{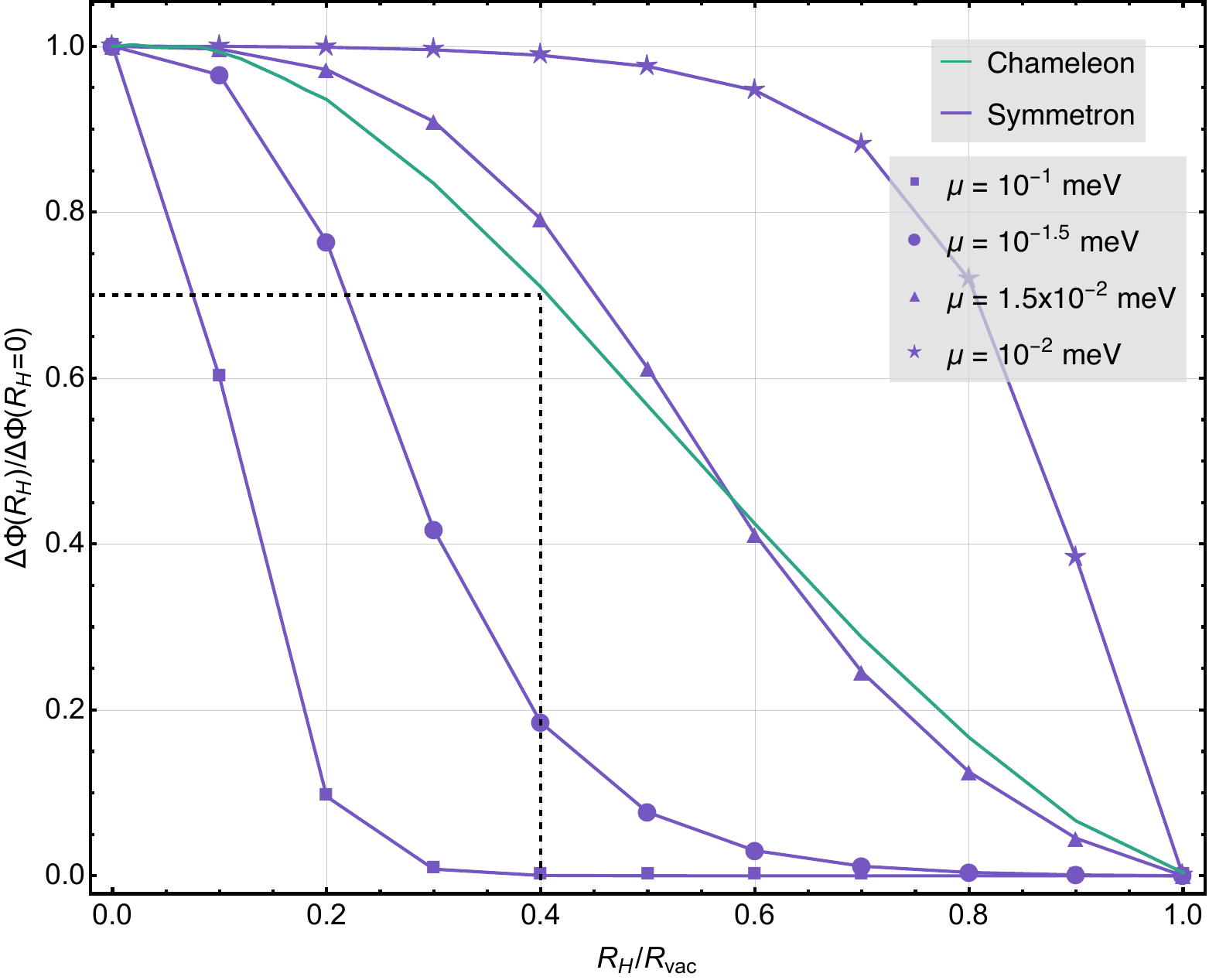}
    \caption{Ratio of gradiometer phase with a central hole of radius $R_{\rm H}$ to that with no hole, $\Delta\Phi(R_{\rm H})/\Delta\Phi(R_{\rm H}=0)$, as a function of $R_{\rm H}/R_{\rm vac}$. The chameleon  (green line) is largely insensitive to $M$ and $\Lambda$ over the relevant parameter range, while the symmetron (purple lines) depends strongly on~$\mu$,  with curves shown for $\mu$ between $10^{-2}\,\mathrm{meV}$ and $10^{-1}\,\mathrm{meV}$. The black dashed line indicates $R_{\rm H} = 0.4 R_{\rm vac}$, which minimally affects chameleon sensitivity while maintaining acceptable signal for $\mu\lesssim10^{-1}\,\mathrm{meV}$ symmetron models.
    }
    \label{fig:hole}
\end{figure}

Figure~\ref{fig:hole} shows the ratio of the gradiometer phase from a plate with a hole to one without, $\Delta\Phi(R_{\rm{H}})/\Delta\Phi(R_{\rm{H}}=0)$, as a function of $R_{\rm{H}}/R_{\rm{vac}}$.
For the chameleon (green line), the relationship between the plate-hole radius and phase is independent of the parameters $\Lambda$ and $M$ within the ranges most relevant to the experiment: $\Lambda = 10^{-4}~\mathrm{meV}-10^{2} ~\mathrm{meV}$ and $M = 10^{-6}~M_\mathrm{Pl}-10^{2}~M_\mathrm{Pl}$. In contrast,  the symmetron (purple lines) result depends strongly on $\mu$, with results shown for $\mu=10^{-2}\,\rm{meV}$ to $10^{-1}\,\rm{meV}$.

The insensitivity to $M$ in the chameleon model is straightforward to understand: for the values of $M, \Lambda$ in question,
the density term $\rho_\mathrm{m}/M$ in the chameleon equation of motion is negligible compared to the $\Lambda^5/\varphi^2$ term, such that the scalar field $\varphi$ is independent of $M$.

Therefore, the fifth force on the atoms scales as $M^{-1}$ and the ratio of the force between the two cases (with and without the hole) remains constant. The lack of $\Lambda$ dependence is more subtle  but may be understood by examining the chameleon field's equation of motion.  As already noted, the density term in the equation of motion is assumed to be negligible for this range of parameters inside the vacuum chamber, and in this limit the equation of motion is invariant under the simultaneous scaling by a constant factor $a$: $\varphi \to a \varphi, \, \Lambda \to a^{3/5} \Lambda$.
For a given value of $\Lambda$, the solution to the problem with a different value is a simple rescaling.  Consequently the force on a test particle also scales as $F \to a F$, as does the integrated phase.  Since the curve in Fig.~\ref{fig:hole} represents a ratio of two integrated phases it is invariant under this scaling.
The relationship between hole diameter and integrated phase is approximately linear across a wide range of $M, \Lambda$ such that a hole of radius $R_H = 0.4 R_{\rm vac}$ results in a phase reduction of approximately $30\%$ compared to a plate with no hole.

A similar argument does not apply to the symmetron.
While it is tempting to try an analogous scaling involving $\mu \to a \mu$, making the equation of motion invariant under this scaling would then require $\varphi \to a^{-2} \varphi$, $\lambda \to a^6 \lambda$, and $\vec x \to a^{-1} \vec x$.  This last transformation on the coordinates~$\vec x$ also scales the boundary conditions, and hence the scaled solution is no longer applicable to our problem.  As such, we simply re-compute the field for each~$\mu$ value of interest.\footnote{Although $\lambda$ dependence was not considered, in fact there is a scaling symmetry $\varphi \to a \varphi,\, \lambda \to a^{-2} \lambda$.  This rescaling leaves $F_\mathrm{hole} / F_\mathrm{solid}$ invariant, so the same curves in Fig.~\ref{fig:hole} are valid for all values of $\lambda$.}

}

For $\mu\geq 10^{-1}$~meV, the phase sensitivity curves  drop off with increasing hole radius more rapidly than for the chameleon because the minimum straight-line distance between the atoms and any portion of the plate increases beyond a Compton wavelength of the scalar field. However, at smaller values, the sensitivity drops more slowly with increasing radius. This may be understood via the reasoning described in Sec.~\ref{sec:model}: if the Compton wavelength is greater than the size of the hole, then the scalar field value is strongly suppressed, such that $\varphi \approx 0$ inside the hole.  The net effect is that the field profile sourced by the plate is similar to that of a solid plate with no hole.

To summarise, the impact of the central hole is modest for the chameleon parameters of interest, while the symmetron sensitivity depends strongly on~$\mu$ and is substantially reduced in the large~$\mu$ limit. This highlights that the optimal hole radius depends not only on the vacuum chamber size and laser optics, but also on the targeted screened scalar theory. 
Our choice of $R_{\rm{H}}=0.4 R_{\rm{vac}}$ represents a balance between signal strength and avoiding laser diffraction and aberration, preserving the chameleon signal whilst maintaining sensitivity for symmetron models with $\mu\lesssim 10^{-1}\,\mathrm{meV}$. Optimising the hole radius for specific theoretical targets, such as larger-$\mu$ symmetrons, would require further study beyond the scope of this work.

\section{Projected Sensitivity}
\label{sec:projections}

\begin{table}[!t]
    \caption{
    Parameters for three potential realisations of a representative 10\,m atom gradiometer experiment. $T_{\rm int}$ denotes the total integration time for the primary measurement campaign at $d_c=1\,\mathrm{cm}$, which excludes calibration runs to parametrise the plate's gravitational field.}
    \begin{tabular}{c c} 
    \toprule
     AI-10\,m & $T_{\rm int}$ \\ 
    \midrule
    Brief &  $400$~s  \\
    Standard &  $12$~hrs \\
    Extended\; &  \;$1$~week  \\
    \bottomrule
    \end{tabular}
    \label{tab:params2}
\end{table}

\begin{figure*}[!t]
    \centering
    \includegraphics[width=0.48\linewidth]{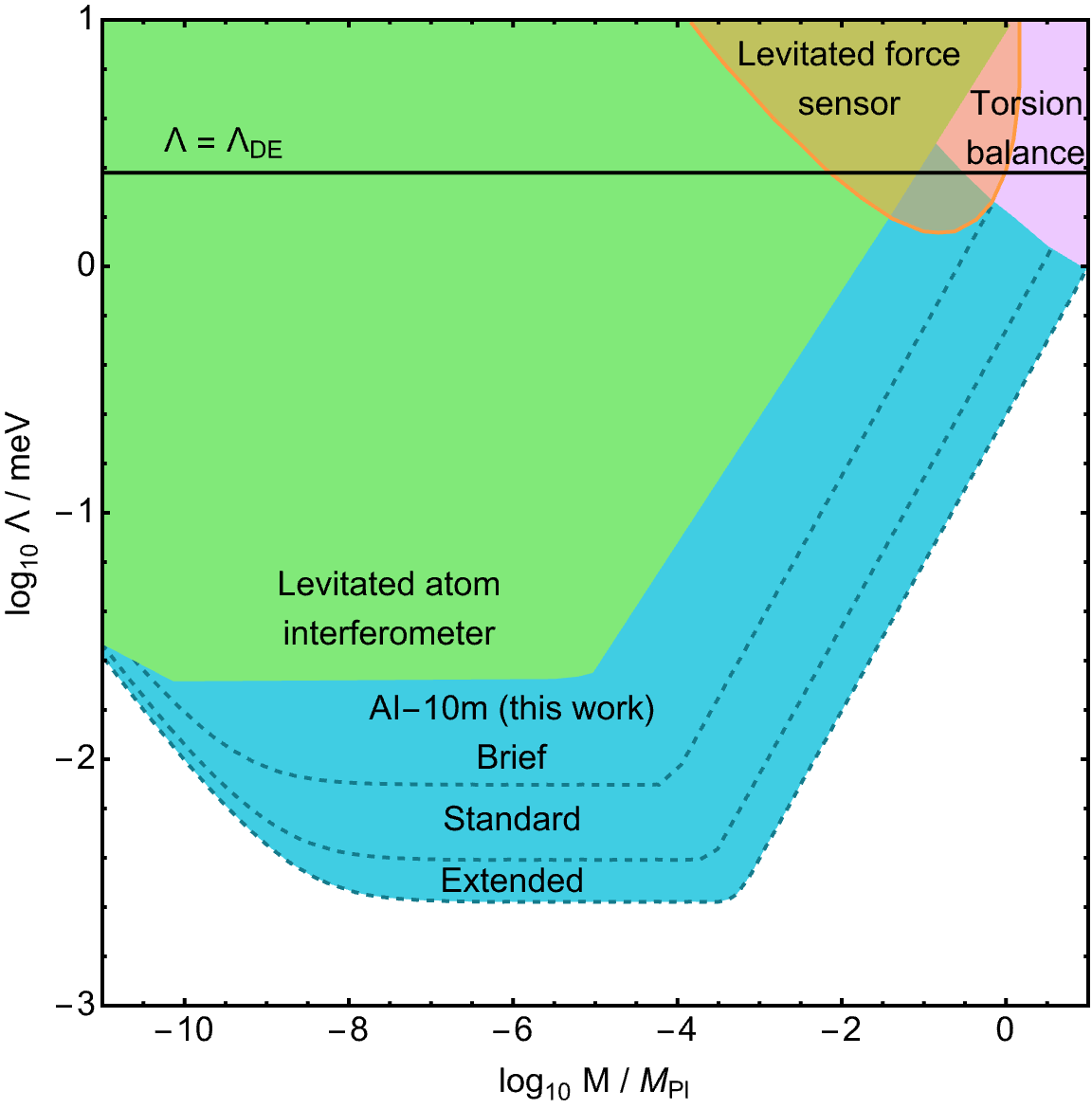}\\
     \includegraphics[width=0.49\linewidth]{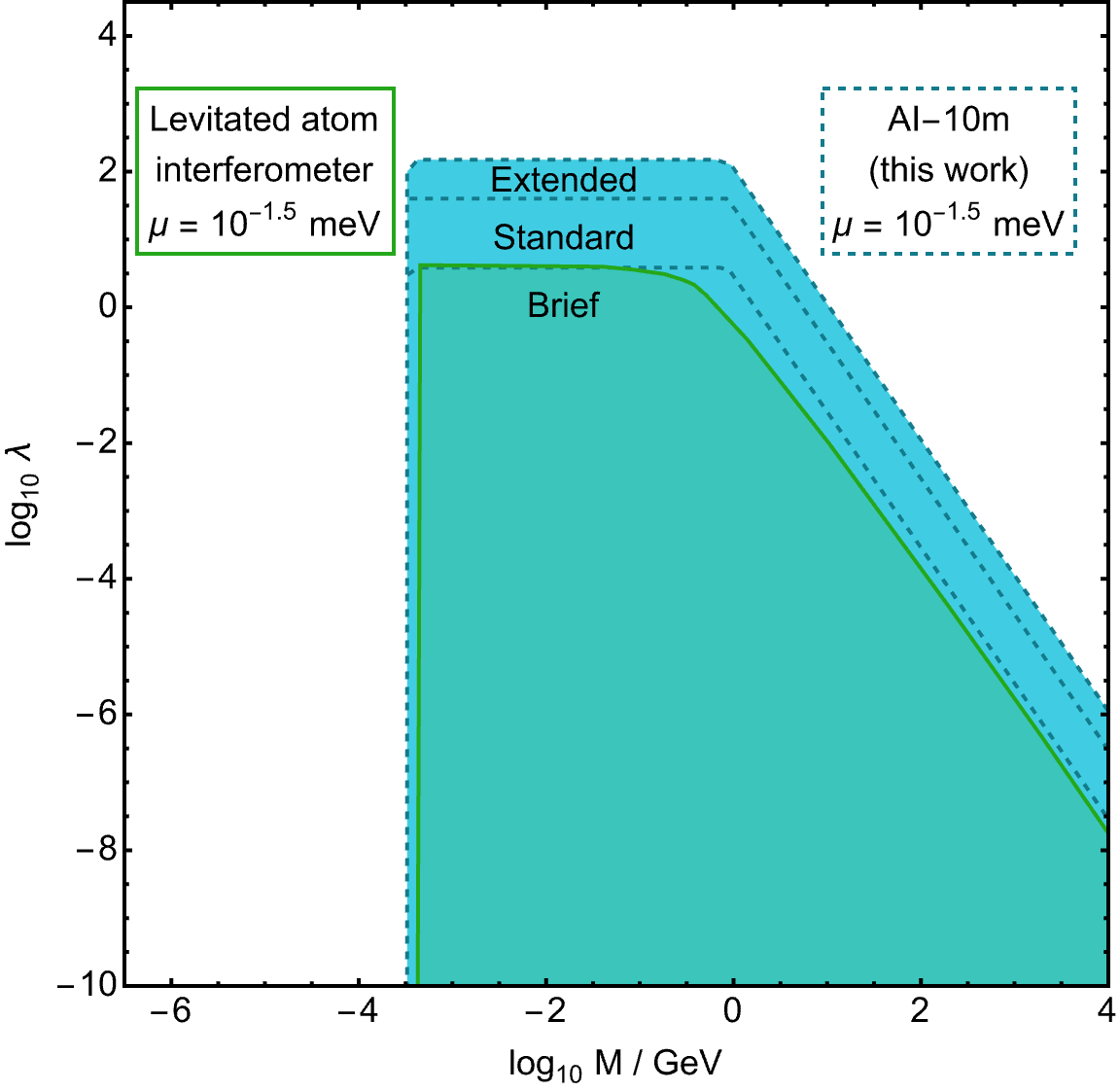}
     \includegraphics[width=0.49\linewidth]{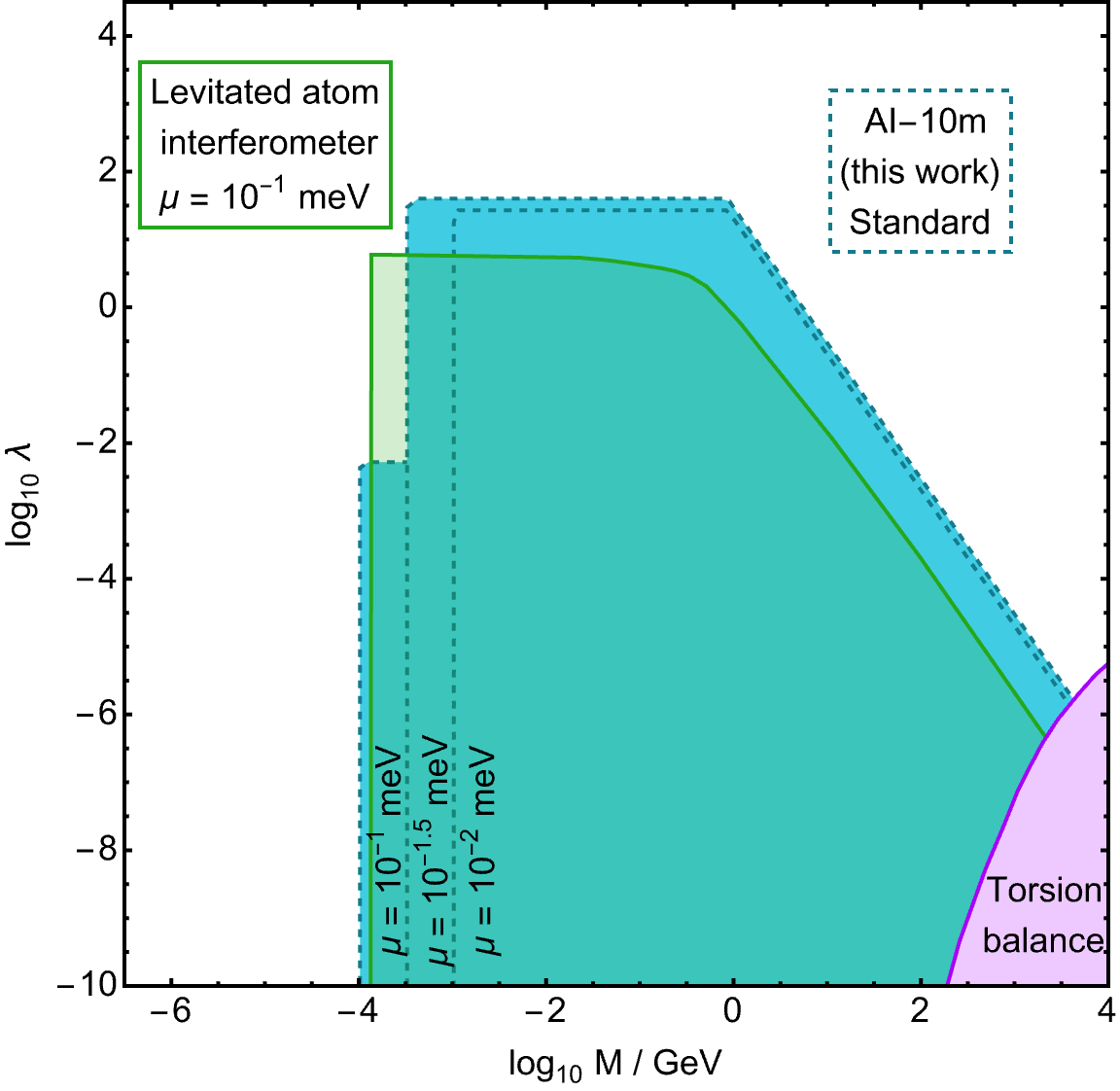}
    \caption{Projected bounds (dashed lines labelled `AI-10m') for a chameleon force (top panel) and symmetron force (bottom panels) with a representative 10\,m atom gradiometer for the Brief, Standard and Extended experimental realisations defined in Table~\ref{tab:params2}.  The bottom left panel shows projections for $\mu=10^{-1.5}$\,meV for all three realisations, whilst the bottom right panel shows different values of~$\mu$ for the Standard realisation only. These are compared to existing experimental constraints (solid lines) from a levitated atom interferometer~\cite{Panda:2023nir}, a levitated force sensor~\cite{Yin:2022geb}, and torsion balance experiments~\cite{Kapner:2006si, Upadhye:2012qu,Upadhye:2012rc}. The torsion balance bound corresponds to $\mu = 10^{-1}~\mathrm{meV}$ and vanishes for smaller values of $\mu$.
    }
    \label{fig:projected_bounds}
\end{figure*}

We now present the projected sensitivity of the proposed experimental protocol. Figure~\ref{fig:projected_bounds} shows the forecasts for chameleon and symmetron theories assuming shot-noise-limited sensitivity. Three potential realisations of this experiment, denoted `Brief', `Standard' and `Extended', are shown, which differ in the duration of the primary measurement campaign at $d_c = 1\,\mathrm{cm}$ as detailed in Table~\ref{tab:params2}. The stated integration times exclude calibration runs to establish the plate's gravitational field at $d_c > R_{\rm{vac}}$, which would require comparable integration times to the primary measurement. The shaded regions indicate parameter space with $\mathrm{SNR} \geq 1$.

The top panel in Fig.~\ref{fig:projected_bounds} shows the sensitivity for a chameleon force in the~$M,\Lambda$ parameter space. The three realisations of the experiment (blue, dashed lines) are projected to surpass previous bounds from levitated atom interferometer experiments (green, solid line)~\cite{Panda:2023nir}, levitated force sensor experiments (orange, solid line)~\cite{Yin:2022geb}, and torsion balance experiments (purple, solid line)~\cite{Kapner:2006si,Upadhye:2012qu}.
These curves exhibit the possibility to robustly test an order of magnitude or more around the dark energy scale $\Lambda = \Lambda_\mathrm{DE}$.

The behaviour of the chameleon sensitivity projections can be understood as follows. At large $M$, the relationship between between $\log_{10}M/M_{\rm Pl}$ and $\log_{10}\Lambda/{\rm meV}$ is linear. Below $M/M_{\rm Pl}\sim10^{-3} \text{ -- } 10^{-4}$, screening of the atom becomes significant: the additional factor of $M$ in the screening factor (as defined in Eq.~\eqref{eq:chameleon-screening-factor}) cancels with the factor of $1/M$ that appears in the acceleration, producing a plateau in the sensitivity projection.
{For example, using Eq.~\eqref{chameleon-sf-atom}, if $\Lambda \approx 10^{-3}~\mathrm{meV}$ the atom is screened for $M \lesssim 10^{-3} M_\mathrm{Pl}$ corresponding to the `knee' on the right side of the figure.} At even smaller $M$, the field reaches the minimum of the effective potential (see Appendix~\ref{app:boundary-conditions}), resulting in a Yukawa-like suppression of the fifth force.

The lower left panel in Fig.~\ref{fig:projected_bounds} shows the forecasted sensitivity of the three experimental realisations to the symmetron for a fixed value of $\mu=10^{-1.5}$~meV.
In this case, our projections for the benchmark setup are comparable to existing bounds for the Brief scenario, with improvements of at least an order of magnitude in $\lambda$ achievable with both the Standard and Extended setup.

{The qualitative characteristics of the lower left panel are as follows.  At large $M$, the atom is unscreened and the constraint follows a power law.  Smaller $M$ corresponds to a stronger fifth force and hence stronger bounds on $\lambda$.  At $M \lesssim \mathrm{GeV}$ the atom becomes unscreened (predicted by Eq.~\eqref{symm-sf-atom}).  From here, increases in the fifth force are offset by decreases in the screening factor, hence the flat portion of the curve.  Moving further to the left, if the critical density of the theory is less than the density of the gas inside the vacuum chamber $\mu^2 M^2 < \rho_\mathrm{vac}$ then the field remains at $\varphi = 0$ everywhere and the fifth force switches off, hence no bounds may be placed.}

The lower right panel of Fig.~\ref{fig:projected_bounds} shows the Standard realisation for different values of $\mu$ spanning $10^{-2}$ to $10^{-1}$~meV. This demonstrates sensitivity improvements over existing bounds, though the magnitude of improvement varies with $\mu$. Similar behaviour is found for the Brief and Extended realisations.

As configured, the proposed setup is most sensitive to $10^{-2}~\mathrm{meV} \lesssim \mu \lesssim 10^{-1}~\mathrm{meV}$, complementing previous atom interferometry tests whose sensitivity peaks at $10^{-1.5}~\mathrm{meV} \lesssim \mu \lesssim 10^{-0.5}~\mathrm{meV}$. The improved sensitivity at lower~$\mu$ values compared to existing experiments results from the larger vacuum chamber radius, as well as the improved phase sensitivity.  The relatively poorer sensitivity at the higher end, $\mu \sim 10^{-1}~\mathrm{meV}$, stems from our choice $d_c = 1~\mathrm{cm}$.  This was chosen conservatively to ensure that systematic effects from the plate, which we discuss in the next section, are small. Reducing $d_c$ may improve the sensitivity to larger values of $\mu$, provided the systematics remain under control.

The chameleon models, on the other hand, have the same Compton wavelength of $m_\mathrm{eff}^{-1} \approx R_\mathrm{vac}$ over a wide range of parameter space and are therefore less sensitive to these considerations.  Hence our forecasts show a more robust improvement for chameleon models, even with our conservative estimate for the plate-atom separation $d_c$.

\section{Systematics}\label{sec:systematics}

The sensitivity forecasts presented in the previous section assume that the experiment is atom-shot-noise limited. Whilst the Newtonian gravitational effect of the plate must be characterised and subtracted from the signal sequence measurements, as discussed in Sec.~\ref{sec:calibration}, other potential sources of systematic uncertainty must also be addressed.  Comprehensive studies of many backgrounds and mitigation procedures have already been discussed in the literature~\cite{MAGIS-100:2021etm,Dimopoulos:2008sv,Dimopoulos:2008hx,Mitchell:2022zbp}, and characterised in 10\,m instruments~\cite{Kovachy2015,PhysRevLett.118.183602,PhysRevLett.111.083001,PhysRevLett.120.183604,PhysRevLett.111.113002}.   In this section we focus on systematic effects particularly relevant to our setup and suggest ways in which they can be suppressed below shot-noise.

A particular advantage of operating in a gradiometer configuration is that systematic effects which are common to the two interferometers, such as laser phase noise, mechanical jitters and uniform magnetic fields, are naturally suppressed in the differential measurement. Of more concern are backgrounds which induce a differential acceleration between the interferometers.
Gravity gradient noise (GGN), which results from fluctuations in the local gravitational field, is one such background. At 10\,m-scales, anthropogenic GGN from moving machinery or personnel is the primary concern~\cite{Carlton:2023ffl}, whilst at longer baselines natural environmental sources may dominate over atom shot noise at sub-Hz frequencies without mitigation~\cite{Badurina:2022ngn, Carlton:2024lqy}.

The fact that the signal is sourced by a controlled object within the laboratory (the plate) provides additional opportunities compared to searches for oscillatory background fields (e.g., ULDM and GW) where the signal frequency is unknown. Here, the signal frequency $f_0$ derives from the $Q$-flip alternation rate (see Fig.~\ref{fig:oscillating_signal}) and is thus a design parameter.  As a result, time-varying backgrounds uncorrelated with $f_0$ can be removed by comparing measurements at different values of $f_0$. In addition, $f_0$ could be selected to avoid the frequency ranges of known background sources.  The choice of $f_0$ is restricted by the duration of each interferometry sequence, although multiplexing atom clouds could alleviate this.

We now consider systematics introduced by the source mass itself.
As discussed in Sec.~\ref{sec:calibration}, the gradiometer phase due to the Newtonian gravity of the plate must be subtracted. Assuming this can be achieved through calibration, the dominant remaining uncertainty arises from shot-to-shot fluctuations in the atom-plate distance~$\Delta z$. This uncertainty induces variations in the gravitational contribution to the gradiometer phase of 
\begin{equation}
    \begin{split}
        \delta\left(\Delta \Phi_{g}\right) = \frac{m_a}{\hbar} \int_0^{2T} {\rm d}t\,\bigg(\big[V_{p,g}(z_2^u+\Delta z) - V_{p,g}(z_1^u+\Delta z)   \big]\\-\big[V_{p,g}(z_2^u) - V_{p,g}(z_1^u)\big]\bigg)~.
    \end{split}
    \label{eq:unc}
\end{equation}
We assume $\Delta z \approx 10\,\mu$m from typical mechanical instabilities in the lab environment such as the uncertainty on the initial position and velocity of the atoms or jitter in the source mass mounting~\cite{Abe:2024idx}.  Using this value in 
Eq.~\eqref{eq:unc} gives a per-shot phase uncertainty of $\delta(\Delta\Phi_g)\sim10^{-5}$~rad. This is below the level of atom shot noise assumed in this work, but may become a limiting factor in the future for experiments with significantly improved shot noise.

Black body radiation from the plate can also impact measurements close to surfaces~\cite{Haslinger_2017}. If the plate temperature is raised relative to the surrounding environment, it will source an additional force proportional to the temperature difference, scaling, like Newtonian gravity, with the inverse squared distance from the plate. This would act differently on the two interferometers and therefore appear in the gradiometer phase. 
With the plate and environment in thermal equilibrium, this effect is expected to be small. Should additional suppression be required, temperature sensors along the baseline would enable modelling and post-correction, whilst random fluctuations could be detrended by exploiting the signal periodicity.

Patch potentials and other short distance forces (e.g. Casimir-Poldor forces) sourced by the plate pose a further potential source of phase uncertainty if the atoms approach too closely. In our sequences, the distance of closest approach of the atoms to the plate is 1\,cm.  At these distances, such effects are expected to be negligible~\cite{garcion2025darkenergysearchatom}.

Finally, we consider systematics arising from the thermal expansion of the atom cloud. 
The impact of cloud dynamics (expansion and temperature) is important for experiments probing short-range effects. In typical experiments, it is expected that the cloud may expand on the order of mm~\cite{Hogan:2008, Badurina:2024nge, Murgui:2025unt}.
Assuming a cloud of diameter $\Delta x = 1$\,mm, the closest and furthest atoms from the plate will experience enhanced and reduced accelerations. Integrating over the sequence gives relative uncertainties of $\Delta\Phi_\varphi{}^{+1.95\%}_{-1.96\%}$. Assuming the cloud is isotropic, we expect these effects to largely cancel out on average, leaving a small relative shift in the measured phase of $\approx -0.01\%$, which we consider negligible.

\section{Discussion \& conclusions}
\label{sec:conclusions}

Screened scalar theories have been proposed as dark energy candidates and arise in broader dark-sector model building. These allow scalars to couple to matter with a strength comparable to, or stronger than, gravity whilst remaining phenomenologically viable through a screening mechanism: in regions of high density, non-linear terms in the scalar equation of motion become large, effectively decoupling the field from the Standard Model. We target two of the most prominent models in this class: the chameleon and the symmetron.

We explored the use of long-baseline atom interferometers to detect or constrain such forces. We proposed an experimental setup employing an annular planar source mass with a central hole ($R_{\rm{H}} = 0.4 R_{\rm{vac}}$) positioned inside the vacuum chamber at the top of a 10\,m-baseline gradiometer. This geometry supplies a scalar field gradient whilst avoiding the strong screening that would occur with an external source mass. The central hole accommodates laser transmission with only a modest $\sim30\%$ signal reduction for chameleon models, whilst retaining good sensitivity for symmetron models with $\mu \lesssim 10^{-1}\, \mathrm{meV}$.

Two key challenges arise with this set-up: first, distinguishing the static fifth force from background systematics; and second, isolating the scalar force from the plate's Newtonian gravity. To address the first, we proposed the `$Q$-flip protocol', which alternates between two distinct sequences. The resonant $Q = 2$ sequence keeps atoms close to the plate throughout their trajectories, whilst the butterfly sequence features crossing arms  causing both the scalar and gravitational contributions to cancel in the gradiometer phase. This alternation induces a periodic time-dependence to the gradiometer phase at a controllable frequency $f_0$, aiding signal extraction and enabling robust noise characterisation through standard time-series analysis techniques. To address the second, we exploit the long baseline which enables measurements beyond the scalar force range ($\sim R_{\rm{vac}}$). This allows in situ gravitational calibration at $d_c \gtrsim R_{\rm{vac}}$ to achieve shot-noise-limited sensitivity when probing close to the plate ($d_c \approx 1\,\mathrm{cm}$).

With these protocols in place, we project sensitivities that lie 1 to 1.5 orders of magnitude beyond existing bounds. For chameleon models, a Standard realisation (12 hours integration time) would decisively test parameter space around the dark energy scale $\Lambda = \Lambda_{\rm{DE}}$, with the Extended configuration (1 week integration time) probing $\Lambda$ from $10^{-2.5}\,\rm{meV}$ to $10^{0.5}\,\mathrm{meV}$ when $M \lesssim M_{\rm{Pl}}$. For symmetron models with $\mu = 10^{-1.5}\,\rm{meV}$, the Standard setup would improve bounds on $\lambda$ by more than an order of magnitude, with greatest sensitivity to $10^{-2}\,\mathrm{meV} \lesssim \mu \lesssim 10^{-1} \,\mathrm{meV}$, complementing previous tests at slightly higher values of $\mu$.

The $Q$-flip protocol offers additional advantages for signal characterisation beyond detection that could be exploited in future work. The power spectral density exhibits peaks at odd multiples of $f_0$, with the pattern of higher harmonics encoding information about the signal. Whilst we focused on the peak at $f_0$ for our sensitivity projections, these features could prove valuable for signal validation in the event of a detection, providing a distinctive signature to distinguish genuine scalar forces from unexpected systematic effects.

The localised nature of screened forces makes 10\,m-scale instruments particularly well-suited for these searches. A key advantage of long baselines is enabling measurements beyond the scalar force range ($\sim R_{\rm{vac}}$) for gravitational calibration, which is fully realised at just a few metres. This contrasts sharply with gravitational wave detection, where  sensitivity scales with baseline and $\sim$km-scale instruments are required. Screened fifth forces thus present an ideal target for the 10\,m atom interferometers that will be operational in the near term, such as AION-10 and VLBAI, with longer 100\,m-scale instruments such as AICE and MAGIS-100 offering further opportunities for complementary searches.

Source mass geometries could be further tailored to optimise sensitivity to screened scalars~\cite{Briddon:2023ayq}. Smaller hole radii could enhance sensitivity to high-$\mu$ symmetrons at the cost of increased laser diffraction and aberration. To eliminate the hole entirely, a solid plate could be moved out of the laser path when pulses are fired and repositioned when the atoms reach the apex, either through horizontal translation or a mechanical shutter mechanism. Whilst this would recover the $\sim30\%$ signal loss from the hole, the engineering challenge is likely to be substantial. 

The measurement techniques developed here are readily applicable to a broad range of theoretical models. A comprehensive survey of screening mechanisms beyond chameleon and symmetron models would be valuable. The experimental configuration could be adapted to probe Yukawa-like forces through larger external source masses, or to search for equivalence-principle violation using different materials or atomic species in the upper and lower arms. Longer baseline instruments offer additional geometric possibilities: periodic test masses positioned along the baseline~\cite{Chiow:2018lze,garcion2025darkenergysearchatom}, a source mass tracking the atomic trajectories to increase interaction time, or multi-gradiometry configurations for enhanced systematic rejection all warrant investigation.

The protocols introduced here provide new routes to probe screened scalar theories. With several 10\,m instruments under construction, the next few years offer timely opportunities to test theoretically motivated regions of screened-scalar parameter space.
With rapidly developing theoretical ideas, experimental innovation, and facilities, long-baseline atom interferometry is set to play an increasingly important role in the search for physics beyond the Standard Model. Beyond application to dark-sector physics, the protocols proposed in this work may be useful for quantum sensing more broadly. 

\section*{Acknowledgments}
We are  grateful to Leondardo Badurina, Clare Burrage, John Ellis, Jeremiah Mitchell, and Chris Overstreet for comments on an early version of this manuscript. 
We also acknowledge discussions and encouragement from Oliver Buchmueller and the wider AION collaboration, and the Terrestrial Very-Long-Baseline Atom Interferometry (TVLBAI) proto-collaboration.
H.B.\ acknowledges partial support from the Science and Technology Facilities Council (STFC) HEP Theory Consolidated grants ST/T000694/1 and ST/X000664/1, and thanks members of the Cambridge Pheno Working Group for  discussions.
J.C.\ acknowledges support from a King's College London NMES Faculty Studentship and acknowledges partial support from the U.S.\ Department of Energy Office of Science under Contract No.\ DE-FG02-96ER40989.
B.E.\ acknowledges support support from STFC Consolidated grants ST/T000791/1 and ST/X000575/1, as well as Simons Investigator award 690508.
T.H.\ acknowledges support from the STFC grant ST/Y004566/1. 
C.M.\ acknowledges support from the STFC grant ST/T00679X/1.

{\bf Data access statement: }The data supporting the findings of this study are available within the paper. No experimental datasets were generated by this research.

\appendix

\section{Scalar field approximations}
\label{app:boundary-conditions}

Whilst in practice we solve the scalar equation of motion Eq.~\eqref{scalar-eom} numerically, much of the behaviour can largely be understood in terms of a small number of analytic arguments which we provide here to offer insight into the behaviour of these fields.  This is also of use to justify various choices that are made in this work, particularly the boundary conditions imposed to solve for the field profile in the experimental apparatus. 
 
For convenience it is common to define the `effective potential' $V_\mathrm{eff}(\varphi)$ to include contributions from matter interactions:
\begin{equation}
    V_\mathrm{eff}(\varphi) = V(\varphi) + A(\varphi) \rho_\mathrm{m}~.
\end{equation}
It is this potential which governs the dynamics of the scalar field.  In the presence of some ambient field value $\varphi$, perturbations to the field value acquire an `effective mass'
\begin{equation}
    m_\mathrm{eff}^2(\varphi) \equiv \frac{\rm{d}^2}{\rm{d} \varphi^2} V_\mathrm{eff}(\varphi)~.
\end{equation}
The inverse of the effective mass is termed the Compton wavelength, and is a useful quantity for understanding the behaviour of the scalar field.  We are now ready to examine in more detail two of the claims made in the main text.  This information has appeared in the literature before~\cite{Upadhye:2012qu,Upadhye:2012rc,Burrage:2014oza,Hamilton:2015zga,Elder:2016yxm,Jaffe:2016fsh,Brax:2018zfb} so we briefly summarise the main points.

\subsection*{Claim 1: Range of the scalar interaction}

It was claimed in Sec.~\ref{sec:setup} that for both the chameleon and the symmetron, the range of the scalar-mediated interaction between the atoms and the plate is at most $\approx R_\mathrm{vac}$.  We discuss the chameleon and symmetron in turn.

For the chameleon, we begin by neglecting the residual gas density inside the vacuum chamber $\rho_\mathrm{vac}$.  Including the density term has the general effect of increasing the Compton wavelength, so this suffices to set an upper bound on the interaction.  In a region with zero density, the chameleon field value rolls to a large value, as may be ascertained by examining the effective potential $V_\mathrm{eff}$.  Given infinite space, the field would reach an infinite value.  But (as we shall see later in this Appendix) the field value must be small, near zero, at the edges of the vacuum chamber.  The solution to the equation of motion minimises the energy of the field configuration, which involves both the (spatially integrated) potential energy $V(\varphi)$ and gradient energy $\frac{1}{2}(\vec \nabla \varphi)^2$.  The field thus rolls to a value finite $\varphi_\mathrm{vac}$ that is a balance between these two contributions to the Hamiltonian density.  It has been shown via numerical calculation~\cite{Hamilton:2015zga,Elder:2016yxm,Brax:2018zfb} that the central field value in the vacuum chamber is
\begin{equation}
    \varphi_\mathrm{vac} = \xi (2 \Lambda^5 R_\mathrm{vac}^2)^{1/3}~,
    \label{chameleon-phi-vac}
\end{equation}
where $\xi$ is an $O(1)$ factor that depends on the geometry of the vacuum chamber.  For an infinitely long cylinder, $\xi \approx  0.68$.  This results in a Compton wavelength of $m_\mathrm{eff}^{-1} \approx R_\mathrm{vac}$, as promised.

For the symmetron, similar reasoning as above leads one to again conclude $m_\mathrm{eff}^{-1} \lesssim R_\mathrm{vac}$, with the caveat that for small mass parameters $\mu < R_\mathrm{vac}^{-1}$ it is energetically favourable for the field to remain at $\varphi = 0$ everywhere, switching off the fifth force entirely~\cite{Upadhye:2012rc,Brax:2014zta}.

\subsection*{Claim 2: The boundary condition $\varphi = 0$}

We now justify the boundary condition used in the numerical calculations, in which $\varphi = 0$ at the surface of the source mass and the vacuum chamber walls.  The essential idea is that inside such dense media the field rolls very quickly to the minimum of the effective potential $\varphi_\mathrm{min}$.  In general this depends upon the parameters of the theory. With parameters where this does not occur, macroscopic objects would be unscreened and such an interaction would already be ruled out by existing fifth force tests.  We now show that for the models of interest we will always be able to take $\varphi = 0$ at the boundaries.

This argument hinges upon two ingredients: (i) the field value reached within the walls $\varphi_\mathrm{min, walls}$ is tiny compared to the typical field values inside the vacuum chamber and (ii) the field reaches $\varphi_\mathrm{min}$ within a negligible distance compared to the size of the wall itself.

We first discuss the chameleon.  Inside a region of constant density $\rho$, the field tends towards the minimum of the effective potential
\begin{align} \nonumber
    \varphi_\mathrm{min}(\rho) &= \sqrt{ \frac{M \Lambda^5}{\rho}}~, \\
    m_\mathrm{eff}^2(\rho) &= 2 \Lambda^2 \left(\frac{\rho}{M \Lambda^3}\right)^{3/2}~,
    \label{chameleon-phi-eq}
\end{align}
where we have also given the effective mass of fluctuations about that minimum.
The field rolls towards the minimum of the effective potential $\varphi_\mathrm{min}$ over a distance of roughly one Compton wavelength $m_\mathrm{eff}^{-1}$.  For the chameleon, inside the metal of the source mass and vacuum chamber walls the Compton wavelength is
\begin{align} \nonumber
     m_\mathrm{eff}^{-1} = \left(9 \times 10^{-2}~\mathrm{cm} \right) &\left( \frac{\rho_\mathrm{wall}}{\rho_\mathrm{water}} \right)^{-3/2} \\
     \times &\left( \frac{\Lambda}{\Lambda_\mathrm{DE}} \right)^{7/2} \left( \frac{M}{M_\mathrm{Pl}}\right)^{3/2} ~,
 \end{align}
where $\Lambda_\mathrm{DE} = 2 \times 10^{-3}~\mathrm{meV}$.  This distance is much less than the typical size of the objects themselves, and of the distance between the atoms and the source mass or vacuum chamber walls (which is of order $\sim \mathrm{cm}$).  We are always interested in the case $\Lambda \lesssim \Lambda_\mathrm{DE}$ and $M \lesssim M_\mathrm{Pl}$, as seen in Fig.~\ref{fig:projected_bounds}.  Likewise, typically $\rho_\mathrm{wall} / \rho_\mathrm{water}$ ranges from roughly 1 to 10 depending on the materials used.  We are therefore justified in setting the field to its equilibrium value $\varphi = \varphi_\mathrm{min}$ at the surface of the source mass and vacuum chamber walls.

We can justify setting $\varphi$ to zero (instead of to $\varphi_\mathrm{eq}$) in a similar manner.  The relevant comparison is between the equilibrium field value in the source mass and walls $\varphi_\mathrm{min, walls}$ and the typical field value in the vicinity of the atoms ($\varphi_\mathrm{vac}$).  Using Eqs.~\eqref{chameleon-phi-vac} and \eqref{chameleon-phi-eq}, we find
\begin{align}\nonumber
    \frac{\varphi_\mathrm{min, walls}}{\varphi_\mathrm{vac}} = 9.7 \times 10^{-2} 
    &\left( \frac{\rho_\mathrm{wall}}{\rho_\mathrm{water}} \right)^{-1/2} \left( \frac{R_\mathrm{vac}}{\mathrm{cm}} \right)^{-2/3} \\
    \times &\left( \frac{\Lambda}{\Lambda_\mathrm{DE}} \right)^{5/6} \left( \frac{M}{M_\mathrm{Pl}}\right)^{1/2}~.
\end{align}
Once again, because we are always interested in the case $\Lambda \lesssim \Lambda_\mathrm{DE}$ and $M \lesssim M_\mathrm{Pl}$, we have $\varphi_\mathrm{min, walls} \ll \varphi_\mathrm{vac}$ so we may set $\varphi_\mathrm{min,walls} \approx 0$.  We are left with $\varphi = 0$ at the surface of the source mass plate and vacuum chamber walls as a reasonable approximation for the real behaviour of the field.

The justification is similar for the symmetron.  Here, the equilibrium field value in the dense material is 0, so it is then only necessary to show that the Compton wavelength is small compared to the size of the objects and the distance to the atoms.  Examining Eq.~\eqref{eq:sym_potentials}, the effective mass of symmetron fluctuations is $m_\mathrm{eff}^2 = - \mu^2 + \rho_\mathrm{wall} / M^2 \simeq \rho_\mathrm{wall} / M^2$.  The Compton wavelength is thus
\begin{equation}
    m_\mathrm{eff}^{-1} = \left(9.4 \times 10^{-6}~\mathrm{cm} \right) \left( \frac{\rho_\mathrm{wall}}{\rho_\mathrm{water}} \right)^{-1/2} \left( \frac{M}{\mathrm{GeV}} \right)~.
\end{equation}
We can now see that this distance is sufficiently small ($\ll \mathrm{cm}$) throughout the parameter space of interest.

\bibliography{refs.bib}

\end{document}